\documentclass{article}
\usepackage[hidelinks]{hyperref}
\usepackage{amsmath}
\usepackage{parskip} 
\usepackage{natbib}
\usepackage{mathrsfs}
\usepackage{pdfpages}
\usepackage{bm}
\usepackage{booktabs}
\usepackage{graphics}
\usepackage{multicol}
\usepackage[bf]{caption}
\usepackage{color}
\usepackage{amssymb}
\usepackage{multirow}
\usepackage{booktabs}
\usepackage{subcaption}
\usepackage{enumerate}
\usepackage{rotating}

\usepackage{comment}
\usepackage{amsthm}

\usepackage[margin=1in]{geometry}
\usepackage{floatrow}
\newfloatcommand{capbtabbox}{table}[][\FBwidth]
\usepackage{rotating}
\usepackage[titletoc]{appendix}
\usepackage[graphicx]{realboxes}
\usepackage{blindtext,titlefoot}
\usepackage{setspace}
\usepackage{color}
\usepackage{lineno}
\usepackage{authblk}

\newcounter{countit}

\title{
Uniaxial compression of 3D printed samples with voids: laboratory measurements compared with predictions from Effective Medium Theory
}

\date{}

\author[1]{\normalsize Filip P. Adamus}
\author[1]{\normalsize Ashley Stanton-Yonge}
\author[1]{\normalsize Thomas M. Mitchell}
\author[2]{\normalsize David Healy}
\author[1]{\normalsize Philip G. Meredith}

\affil[1]{\footnotesize {Department of Earth Sciences, University College London, London, UK}} 
\affil[2]{\footnotesize {School of Geosciences, University of Aberdeen, Aberdeen, UK \\ \linebreak
\[adamusfp@gmail.com \quad ashley.sesnic.18@ucl.ac.uk \quad d.healy@abdn.ac.uk  \quad tom.mitchell@ucl.ac.uk \quad p.meredith@ucl.ac.uk\]}}

\begin{document}
\maketitle
\begin{abstract}
3D printing technology offers the possibility of producing synthetic samples with accurately defined microstructures. As indicated by effective medium theory (EMT), the shapes, orientations, and sizes of voids significantly affect the overall elastic response of a solid body. By performing uniaxial compression tests on twenty types of 3D-printed samples containing voids of different geometries, we examine whether the measured effective elasticities are accurately predicted by EMT. To manufacture the sample, we selected printers that use different technologies; fused deposition modelling (FDM), and stereolithography (SLA). 
We show how printer settings (FDM case) or sample cure time (SLA case) affect the measured properties. We also examine the reproducibility of elasticity tests on identically designed samples. To obtain the range of theoretical predictions, we assume either uniform strain or uniform stress. Our study of over two hundred samples shows that measured effective elastic moduli can fit EMT predictions with an error of less than 5\% using both FDM and SLA methods if certain printing specifications and sample design considerations are taken into account. Notably, we find that the pore volume fraction of the designed samples should be above $\approx1\%$ to induce a measurable softening effect, but below $\approx5\%$ to produce accurate EMT estimations that fit the measured elastic properties of the samples. Our results highlight both the strengths of EMT for predicting the effective properties of solids with low pore fraction volume microstructural configurations, and the limitations for high porosity microstructures. 

\end{abstract}
\section{Introduction}
The elasticity of a material containing voids (empty pores) or inclusions (fluid-filled or solid pores) can be approximated using effective medium theory (EMT). The theory allows one to homogenise a medium and view its response on a larger scale. In turn, understanding how the microstructure (pores shape, orientation, and size) influences the overall response is beneficial in many disciplines, such as materials science, biomechanics, rock mechanics, seismology, oil and gas exploration, among others.

Probably the first steps in treating a material as effectively elastic were done by \cite{Voigt89}, \cite{Reuss29}, and \cite{Bruggeman37}. 
Their works were followed and greatly developed by~\cite{Hill52},~\cite{Eshelby57},~\cite{Hashin59}, and \cite{Bristow60}; researchers inspired by problems in metallurgy.
Subsequently, their averaging approach was expanded by~\citet{Walsh65},~\citet{Budiansky65},~\citet{Hill65},~\citet{OConnell74},~\citet{Kachanov80},~\citet{OdaEtAl84},~\citet{Zimmerman86}, and~\citet{Sayers91} who described how the specific geometry of ellipsoidal pores and cracks affect the stiffness (compliance) of a solid; a rock in particular. 
Independent efforts were made in the field of seismology by~\citet{Postma55},~\citet{Backus62},~\citet{Sch88}, and~\citet{Sch97}, who focused on layered inhomogeneities or long and thin cracks instead of ellipsoidal shapes. Nevertheless, their approaches give approximately equal predictions when the layered features are viewed as a series of aligned penny-shaped cracks~\citep{Sch88}.  All effective elasticity approximations utilised by the aforementioned authors---and numerous researchers not mentioned herein---share the same concept, that the elasticity of the constituents (i.e., solid matrix and inhomogeneities) can be averaged to predict the overall medium's stiffness under given boundary conditions. Because there is no single recipe on how to do it, authors often use the notion of ''effective medium theory'' to refer to this general concept of averaging. The EMT goes beyond elasticity, with its origins in electromagnetic properties of materials; it makes an essential part of a multidisciplinary field called micromechanics~\citep{Kach18}. 

The EMT prediction for even the simple scenario of a single void embedded in a homogeneous matrix is not straightforward. The approximation is unequivocal only if 1) the background is isotropic or transversely isotropic, 2) the void is ellipsoidal, and 3) the void is very small. In the case that one of the first two conditions is not satisfied, an analytical solution cannot be obtained~\citep{Kach18}. 
In the case that the void is relatively large, then the prediction will vary depending on the assumption chosen, due to the boundary effects~\citep{Hill52}. Then, the so-called apparent, and not effective, elasticity parameters are considered~\citep{Huet90}. One possibility is to assume uniform strain throughout the sample. Another is to assume uniform stress. Either assumption leads to an approximately identical result if the inhomogeneity is small, but are discrepant for larger sizes.
This discrepancy also arises if there are more than one void that interact with each other, thus disturbing the stress and strain fields. If the distribution of voids is sparse, then the so-called non-interactive approximations are proposed~\citep{Kach92}. If voids are close together or even overlapping slightly, the interaction effect needs to be taken into consideration~\citep{Hill65}. 

Early studies on ceramics with spherical pores have confirmed empirically the usefulness of EMT~\citep{Walsh65}. However, empirical validation becomes more difficult in the case of strongly oblate or prolate spheroids, shapes that may be good approximations for cracks present in e.g., rocks. Because geomaterials usually have complex microstructures, the empirical validation of EMT using rocks is limited. For instance, we can verify that cracks of known orientation reduce the magnitude of specific parameters in the elasticity tensor~\citep{BrantutPetit22}, and therefore weaken the rocks, as predicted by the theory. However, the magnitude of the effect remains difficult to evaluate due to uncertainties in crack geometries, crack concentration, and crack interactions.

There have been attempts to examine EMT on synthetic materials, where the microstructure is much better constrained. One approach is to place polymeric inhomogeneities inside a previously prepared sandy matrix,with the polymeric materials being subsequently removed by chemical or thermal processes, leaving voids behind~\citep{SantosEtAl17}.
Another method involves rubber pieces placed in epoxy glass~\citep{HenriquesEtAl18}. Alternatively, one may try to use laser etching to create cracks in glass~\citep{StewartEtAl13}. Finally and more recently, researchers used 3D printing technology to create synthetic samples with pores~\citep{HuangEtAl16,zerhouni2019numerically}. 

To our knowledge, researchers have used either manufacturing or 3D printing samples having small inhomogeneities only---such as highly concentrated small pores \citep{NdaoEtAl17, Zerhouni19, DugarovEtAl22}.
Pores of a much smaller size than the wavelength are required for acoustic measurements in order to satisfy the quasi-static assumption of EMT. This way, a full elasticity tensor can be obtained at the expense of inaccuracy in the manufacture of small pores. Besides the inaccuracy in microstructure design, small pores do not soften the material significantly and limit the possibility of interaction validation. In other words, small pores must be densely packed in order to obtain a measurable weakening effect on stiffnesses. In turn, one cannot clearly distinguish the impact of interactions from the effect of pore concentration itself. Also, the small size of pores impedes fluid injection to measure the poroelastic response of such samples. All the aforementioned problems can be overcome if larger inhomogeneities are considered. A downside of such a solution, though, is that the acoustic measurements are heavily degraded due to the static assumption violation. Also, the theoretical prediction is not unequivocal due to the impact of the choice of boundary conditions. 

In this paper, we used fused deposition modelling (FDM) and stereolithography (SLA), two standard 3D printing technologies, to produce twenty different type of samples that contain relatively large voids with different geometries. 
Our samples are uniaxially compressed, and they behave elastically within a significant range of deformation.  
We obtain the elastic components of the stiffness matrix (Young's modulus and Poisson's ratio) from the experimental measurements, and attempt to validate EMT empirically. 
Pores are designed to be large enough to highlight the softening effect and ensure that the impact of void interactions is pronounced. 
Also, the sizes of voids might facilitate future fluid injection tests.
3D printing provides a very high accuracy of the desired microstructure that may overcome the above-mentioned manufacturing methodologies. 
Although 3D printing was developed in the 1980s, to our knowledge, only a few works have used this technology in the context of effective elasticity measurements~\citep{HuangEtAl16, NdaoEtAl17, Zerhouni19, DugarovEtAl22}. 

Our research fills the gap in EMT validation studies for a variety of reasons. First, different geometries of voids are designed---instead of dense microcracks only---that are large enough to enhance the softening effect and overcome measurement errors. Also, we propose various interaction configurations that have not been previously examined experimentally. Further, we performed novel experiments where cylindrical samples containing twenty different void microstructures, printed with two different technologies, were deformed under uniaxial compression.  
Finally, we report results from over two hundred samples in total, which means that our experimental validation is less prone to random errors.
\section{Main problems addressed}
In this paper, we discuss three main issues regarding uniaxial compression tests of 3D-printed samples with voids. First, we attempt to understand how the differences between 3D printing technologies and their specifications may affect the effective Young's modulus measurements. Second, we present the impact of Young's modulus measurement methodology. Third, and most importantly, we focus on whether EMT accurately predicts Young's modulus given the issues mentioned above. 
\subsection{3D printing technologies}
3D printing has revolutionized the manufacturing industry, and numerous different printing technologies have emerged during the past decades, including Stereolithography (SLA), Selective Laser Sintering, Fused Deposition Modelling (FDM), Digital Light Process, and PolyJet, among others. These technologies range in price, accessibility, post-printing processing requirements, printing resolution, and surface finishing quality. Two standard 3D printing technologies were used in this study: SLA and FDM.

Stereolithography, patented in 1986, is the world's pioneering 3D printing technology. This method is a type of additive manufacturing process that employs a laser source to solidify liquid-state, light-reactive resin into solid polymers through a process called photopolymerization. A Computer-Aided Design (CAD) model is imported into the 3D printer which then constructs the object layer by layer (with $0.1$ mm resolution). This is achieved by focusing the laser source,  by which it selectively solidifies the liquid resin, precisely following the cross-sectional images of the 3D model. The resulting products usually require post-printing treatment. They are first immersed in isopropyl alcohol (IPA) to remove any residual, uncured resin. Subsequently, an additional curing step under heat may be needed to finalize the polymerization reaction, which brings the material to its optimal mechanical properties. The SLA method is known for producing highly detailed and accurate objects with seamless surface finishes. This makes it an optimal choice for a wide range of applications, including prototyping, jewelry craftsmanship, and dental modelling, among others. Once the resin is hardened, though, it cannot be recycled. Discarded parts need to be disposed as household waste. 

Fused Deposition Modelling is a widely used, cost-effective 3D printing technology. As in SLA printing, FDM also constructs CAD modelled objects by additive manufacture of the model cross-sections. A thermoplastic filament, in our case, biodegradable, polyactic acid (PLA), is fed into a heated nozzle that melts the material and then precisely extrudes it onto the build platform, constructing the object layer by layer. FDM printing is characterised by the lowest entry and material price of the 3D printing market, by very simple usage, and for having the lowest environmental impact with respect to other 3D printing technologies \citep{li2017cost}. The latter is mainly due to its lower process energy consumption and the possibility of reusing or recycling its wasted products. These factors make FDM a highly popular choice for proof of concept purposes and prototyping. However, the level of accuracy and precision that FDM can offer is limited by the diameter of the circular orifice of the nozzle tip (in our case $0.4$ mm). In addition, the layer-by-layer deposition (with $0.1$ mm of resolution) limits the quality of curved surfaces, causing a grainy surface finishing.  

The ability of defining the exact shape, position, orientation, and density of the pore microstructure in a homogeneous matrix makes 3D printing highly attractive for evaluating the predictive capacity and limitations of EMT. In the context of 3D printing for EMT testing purposes, the starting material needs to be as isotropic, non-porous, and homogeneous as possible. Because both SLA and FDM printed objects are built layer by layer, some anisotropy is to be expected in the mechanical properties of printed materials. Mechanical anisotropy has been shown to be low ($\approx1\%$) for SLA printed objects \citep{kazmer2017three} although it can be affected by the prescribed resolution/layer thickness \citep{chockalingam2006influence}. A helium pycnometer was used to measured the connected porosity of our 3D printed cylindrical samples, obtaining a porosity of $1.2\%$ for whole SLA printed samples.

By contrast, mechanical anisotropy can be significant for FDM printed samples ~\citep{kazmer2017three}. 
The magnitude of mechanical anisotropy is affected by several printing parameters, which define the drawing pattern to be followed by the nozzle tip. 
One or more perimeters are first drawn to delimit the object's boundary. Then, the nozzle tip moves back and forth at a prescribed angle (raster angle) to fill the delimited perimeter(s). The prescribed amount of delimiting perimeters, the raster angle, and the layered structured of the samples will determine the mechanical anisotropy of FDM printed samples~\citep{mohamed2015optimization}. To verify the magnitude of anisotropy of FDM printed samples, we performed acoustic velocity measurements on four PLA cubes ($30$ mm$^3$) with two perimeters, rectilinear infill---the only recommended pattern for 100\% infill---and default other printing specifications. 
Our results indicated $0.8\%-3.3\%$ of anisotropy in the direction orthogonal to layering compared with parallel, which can be considered as a satisfactory in the context of EMT predictions. 
Another potential issue is caused by the air gaps between layers and rasters that result in an intrinsic porosity for FDM 3D printed samples of around $5\%$, as measured by a helium displacement pycnometer.
Main aforementioned features of both 3D printing technologies are listed in Table~\ref{tab:list}.
\begin{table}[!htbp]
\scalebox{1}{
\begin{tabular}
{llllc}
 Feature & SLA & FDM  & in favour of\\
\toprule
Resolution and accuracy & higher & lower & SLA\\
Surface finishing& smooth & not smooth& SLA \\
Mechanical anisotropy & lower & higher & SLA \\
Porosity & 1\% &5\% & SLA \\
Post-printing processing & required & not required & FDM\\
Printer and material cost & higher & lower & FDM\\ 
Environmental impact & higher & lower & FDM\\ 
\bottomrule
\end{tabular}
}
\caption{\footnotesize{Comparison of features between SLA and FDM 3D printing technologies in the context of EMT testing }}
\label{tab:list}
\end{table}
\subsection{Young's modulus measurements} \label{moduli}
Young's modulus is defined as the slope of the elastic (straight-line) portion of a stress-strain curve. 
If the slope is taken between two stress-strain points, the modulus is the change in stress divided by the change in strain.
In general, brittle materials such as metals, certain plastics, and rocks will exhibit steeper slopes and higher moduli values than ductile materials such as rubber. 
Importantly, the values of Young's moduli may differ depending on the derivation methodology.

\citet{DeanEtAl95} distinguish at least three terms used to specify the Young's modulus value; tangent modulus, chord modulus, and secant modulus. 
The first term describes a modulus obtained from a tangent line to the stress-strain curve. According to the ASTM standards for compression measurements of rigid plastic materials (ASTM-D695), a tangent line should be constructed at any point in the straight-line part, whereas if there is no clearly linear region, one should construct a tangent to reach
the maximum (steepest) slope of the curve.
The second term describes a modulus measured from the selected start and end points of the curve. A line is drawn between these two points and the slope of that line is recorded as the modulus. Finally, the secant modulus is a particular type of chord modulus, where the start point is defined as the origin of the stress/strain curve.

As reported by \citet{MalkowskiEtAl18}, based on over two hundred compressed rock samples, Young's moduli values differ significantly based on the choice of the start and end points for the modulus calculation. They specified these points based on: 1) the lower and upper elastic limits of the curve (if an elastic part is perfectly linear, this approach is equivalent to the ASTM-D695 tangent method), 2) 20\% and 80\% of the ultimate strength (cord modulus), and 3) 0\% and 50\% of the ultimate strength (secant modulus). Their analysis indicates a large discrepancy in results between the secant modulus and the two former methods (circa 23\%) due to the inelastic crack compaction occurring in the initial part of the curve. 
\citet{MalkowskiEtAl18} suggest that, in the case of sedimentary rocks, the secant modulus measure is inadequate, whereas chord modulus of a narrower range (e.g., 30\%--70\% of ultimate strength) might be the most efficient measure of Young's moduli.

In this paper, our primary goal is to consider the impacts of the voids on the elastic moduli of materials. These impacts might be different depending on the Young's moduli derivation methodology chosen. Therefore, similarly to~\citet{MalkowskiEtAl18}, we compare different Young's moduli measures but on plastic samples (see Section~\ref{sec:mod}). The comparison itself thus becomes an important part of the study.

\subsection{Effective medium predictions}\label{sec:emt}
In this section, we discuss the elastic moduli that we measure during uniaxial tests, the properties that are required to obtain the theoretical predictions, and we indicate how these predictions are calculated. 

In a uniaxial compression test, the components of the principal stress tensor can be written as $\sigma_{1}\neq0$, $\,\sigma_2=0$, and $\,\sigma_3=0$.
In every test, we measure the axial strain in the direction of compression that---invoking Hooke's law--- is equal to $\varepsilon_1=S_{1111}\sigma_1$, where $S_{ijk\ell}$ denotes the forth-order compliance tensor. 
Additionally, if a sample is isotropic or transversely isotropic (TI) with the symmetry axis aligned with the compression direction, we measure the radial strain that is equal to $\,\varepsilon_r=\varepsilon_2=\varepsilon_3=S_{1122}\sigma_1$.
The above elastic properties can be expressed as Young's modulus and Poisson's ratio, namely, $E:=\sigma_1/\varepsilon_1=1/S_{1111}$ and $\,\nu:=-\varepsilon_r/\varepsilon_1=-S_{1122}/S_{1111}$, respectively.
In the case of a compression test of a plain sample, both Young's modulus and Poisson's ratio are background moduli used in the theoretical predictions, as explained later.
In the case of a sample with a void (that usually induces anisotropy of lower symmetry than TI), we measure Young's modulus only, which is the effective parameter.

Effective properties of a medium with a single inhomogeneity (e.g., void) can be predicted differently depending on the boundary conditions we choose. If stress is assumed to be uniform under the absence of the inhomogeneity, then according to e.g.,~\citet{Shafiro97},
\begin{equation}\label{singlecavity}
S^{\rm{eff}}_{ijk\ell}=S_{ijk\ell}^0+\phi H_{ijk\ell}+\phi \Delta H_{ijk\ell}\,,
\end{equation}
where $S_{ijk\ell}^0$ is the background compliance tensor, $\phi$ is the inhomogeneity volume fraction, $H_{ijk\ell}$ is the compliance increase due to void presence, and $\Delta H_{ijk\ell}$ is the compliance correction factor due to the presence of fluid or solid in the void. 
If displacement, not stress, is assumed to be uniform, then
\begin{equation}\label{singlecavity2}
S_{ijk\ell}^{\rm{eff}}=\left(C_{ijk\ell}^0+\phi N_{ijk\ell}+\phi \Delta N_{ijk\ell}\right)^{-1}\,,
\end{equation}
where $C_{ijk\ell}^0$ is the background stiffness tensor, $N_{ijk\ell}$ is the stiffness decrease due to the presence of the void, and $\Delta N_{ijk\ell}$ is the stiffness correction factor due to the presence of fluid or solid in the void.
In this paper, we do not consider fluid or solid filling of the void; hence, the last term in~(\ref{singlecavity})--(\ref{singlecavity2}) equals zero. 
We use (\ref{singlecavity}) and (\ref{singlecavity2}) to obtain the maximum and minimum predicted value of Young's modulus, respectively. 
Both expressions give approximately equal values if the void is small. The range of prediction rises with larger $\phi$.

$H_{ijk\ell}$ and $N_{ijk\ell}$ can be derived analytically for ellipsoids embedded in an isotropic or TI background. They depend on a) void shape, b) void orientation, and c) background Poisson's ratio.  
The exact expressions for $H_{ijk\ell}$ and $N_{ijk\ell}$ are presented in e.g.,~\citet{Kach18} (Appendix A).
Therefore, the range of effective Young's modulus can be predicted if we know the geometry of the void, and we measured Young's modulus and Poisson's ratio of the plain sample, since the same elastic properties are assumed to describe the background of the sample with a void.

In the case of multiple voids, if one assumes that the distance between inhomogeneities is great enough, then the stress at the boundary of each void region is nearly the same as at the boundary of the body. We can write the so-called non-interaction approximation (NIA) as
\begin{equation}\label{NIA}
S^{\rm{eff}}_{ijk\ell}=S_{ijk\ell}^0+\sum_m\phi^{(m)} H_{ijk\ell}^{(m)}\,,
\end{equation}
where superscript $m$ describes each void in the medium.
If the displacement at the boundary of each void region is nearly the same as at the boundary of the body, we get a dual version of NIA,
\begin{equation}\label{NIA2}
S_{ijk\ell}^{\rm{eff}}=\left(C_{ijk\ell}^0+\sum_m\phi^{(m)} N_{ijk\ell}^{(m)}\right)^{-1}\,.
\end{equation}
In the case where voids are relatively close to each other, the interaction between them may disturb the stress and displacement field. This is why several interaction schemes, such as the self-consistent, differential, Maxwell, or Mori-Tanaka schemes, were proposed~\citep{Kach92}. Nevertheless, all of them use NIA as a building block and predict effective compliances within the range provided by expressions~(\ref{NIA})--(\ref{NIA2}). Also, unless voids are parallel, these schemes are difficult to apply in the cases of anisotropy~\citep{Kach18}.
Further, they are designed to take into account larger concentrations of randomly oriented voids. Since in this paper we consider cases that are both anisotropic and with low concentration of voids, we focus on whether the measured values of Young's moduli fall into the range of NIA dual predictions only; which is sufficient for our purposes. 
\section{Experiments}
\subsection{Experimental setup: uniaxial compression}
\begin{figure}[!htbp]
\centering
\begin{subfigure}{.4\textwidth}
\includegraphics[scale=0.7]{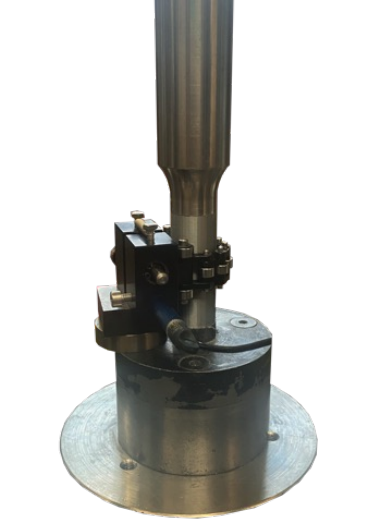}
\caption{}
\label{uniaxial}
\end{subfigure}
\begin{subfigure}{.4\textwidth}
\includegraphics[scale=0.8]{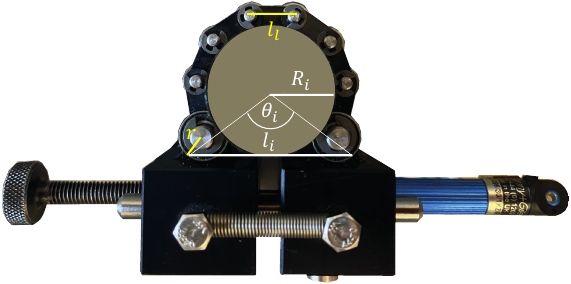}
\caption{}
\label{extensometer}
\end{subfigure}
\caption{\footnotesize{\bf{a)}} 3D printed sample loaded in the uniaxial frame. {\bf{b)}} Extensometer and parameters to obtain circumferential strain.}
\label{Figure1}
\end{figure}

3D printed samples were subjected to uniaxial compression using a servo-controlled uniaxial load frame in the Rock and Ice Physics Laboratory of University College London (Figure \ref{uniaxial}). Samples were loaded at a constant strain rate of $10^{-5}$$s^{-1}$ until yielding. Axial strain was continuously measured using an external linear variable differential transformer (LVDT) displacement transducer, while load was recorded using a 200kN load cell. Corrections were applied to the axial displacement data to account for machine stiffness.
Intact (plain) 3D printed cores were uniaxially deformed while taking continuous measurements of circumferential strain using an LVDT extensometer (Figure \ref{extensometer}). The correction applied to convert from linear displacement (LVDT output) to circumferential strain is detailed as follows. The initial aperture angle $\theta_{i}$ is obtained from the initial sample radius $R_{i}$, the total chain length $l_{c}=7l_{l}$, where $l_{l}$ is the link length, and extensometer rod radius $r$ from,

\begin{equation}
\theta_{i}=2\pi - \frac{l_{c}}{R_{i}+r}.
\end{equation}

The circumferential strain $\epsilon_{c}$ is calculated from the differential LVDT output $\delta l=l_{f}-l_{i}$ (see Figure~\ref{extensometer}) as,
\begin{equation}
\epsilon_{c}=\frac{2\pi\delta l}{\sin(\theta_{i}/2)+\cos(\theta_{i}/2)(\pi-\theta_{i}/2)}.
\end{equation} 

\subsection{Samples: configuration and preparation}
We designed twenty types of cylindrical samples having different configurations of voids. These samples were $62.5$ mm long with a diameter of $25$ mm. This way, the $2.5$ proportion between length and diameter suggested by ASTM-D695 and rock experimentalists e.g.~\citep{PatersonWong05} is achieved.
We divided the samples into two groups designed for two phases of experiments. In the preliminary phase of tests (numbering of samples preceded with 1), we compressed only five types of cores that are either plain or have a single void (samples 1.1--1.5). Their simple configuration allowed us to verify the influence of 3D printer type, specifications (or cure time), and long-crack edge effects on the results (see Section~\ref{sec:edge}). Subsequently, in the second phase, we chose the best performing---in the context of measurement/prediction error---3D printer and specification to perform experiments on fifteen types of cores having voids with diminished edge effect and more complicated configurations (samples 2.1--2.15). Let us discuss the configuration and preparation of the samples from each group separately.

\begin{figure}[!htbp]
\centering
\includegraphics[scale=0.5]{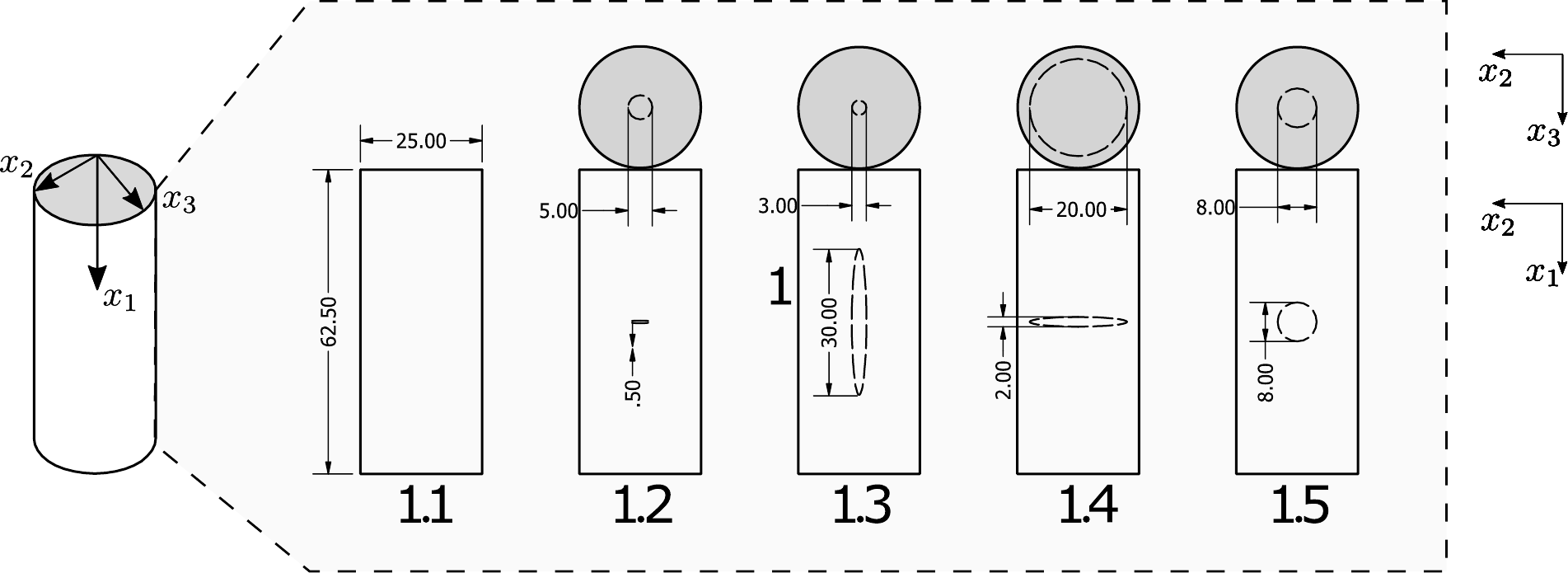}
\caption{\footnotesize{Geometry of samples designed for the preliminary phase of the experiments, where dimensions are in millimetres. Sample 1.1 is plain, whereas samples 1.2 and 1.4 consist of strongly oblate spheres having a $0.1$ aspect ratio. }}
\label{fig:phase_one}
\end{figure}

\begin{figure}[!htbp]
\centering
\includegraphics[scale=0.12]{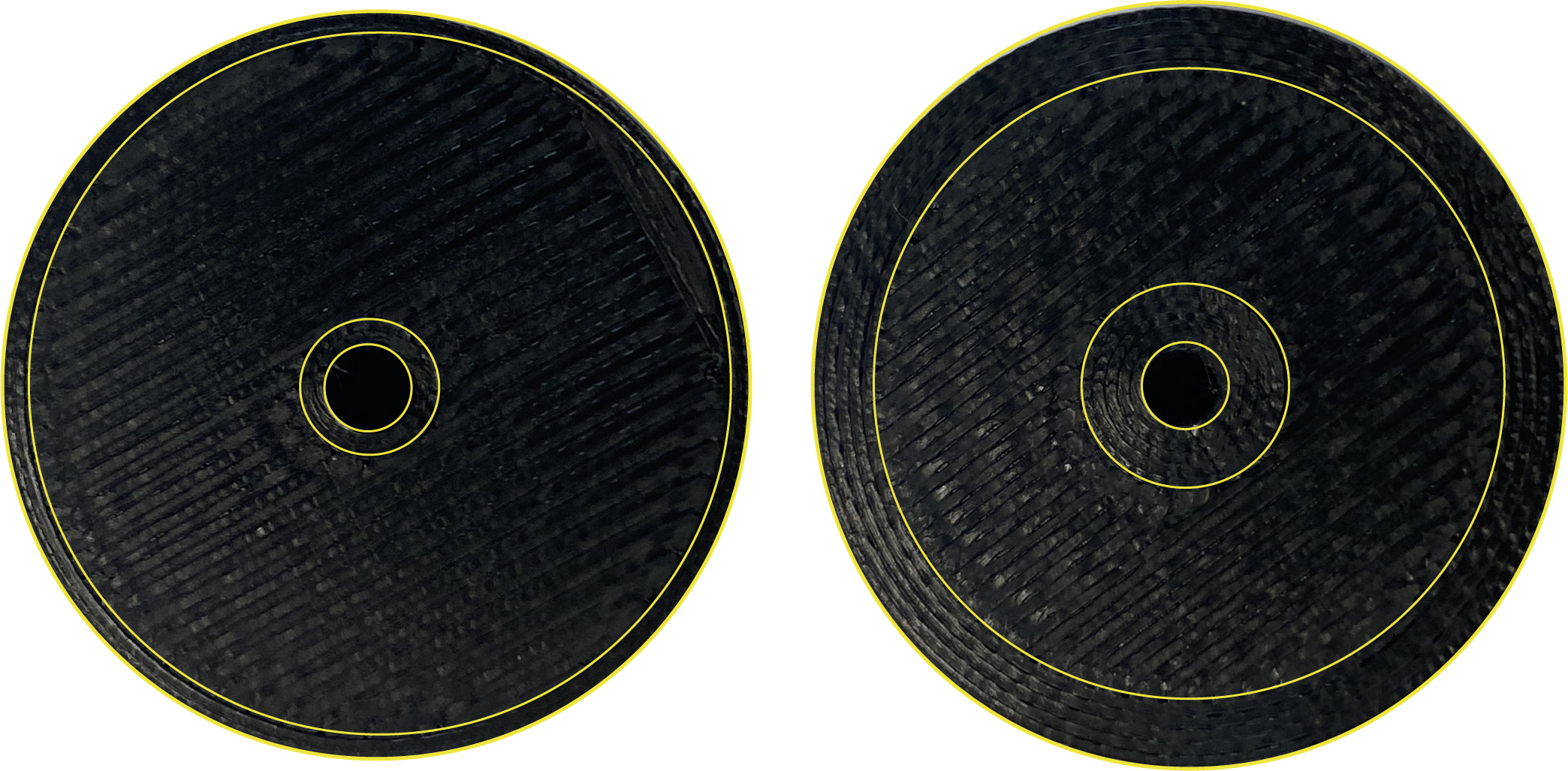}
\caption{\footnotesize{Effect of prescribed perimeters in FDM (two in the left, five to the right).} }
\label{perimeters}
\end{figure}

\begin{figure}[!htbp]
\centering
\includegraphics[scale=0.218]{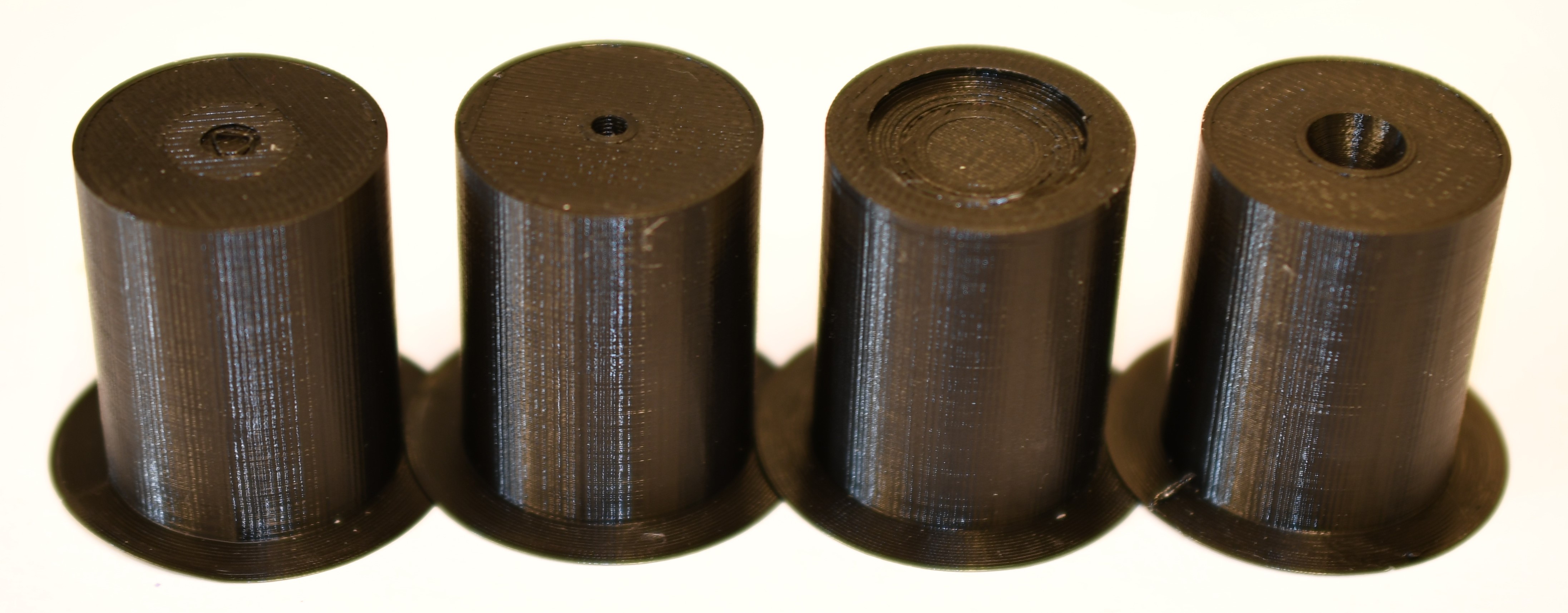}
\caption{\footnotesize{Half-printed FDM samples having two perimeters used in the preliminary phase of experiments. Samples have brims that prevent their detachment from the bed.} }
\label{samples_starters_FDM}
\end{figure}

The longitudinal and cross-sections of samples designed for the preliminary phase of experiments are depicted in Figure~\ref{fig:phase_one}. Concise description of the voids geometries can be found in Table~\ref{tab:phase_one}. Sample 1.1 is necessary to obtain the elastic properties of the background material that are used in the theoretical calculations~(\ref{NIA})--(\ref{NIA2}). Theoretically, a void with a very low aspect ratio makes the softening more pronounced, an this effect should be even more significant if the volume of the pore increases. Therefore, samples 1.2 and 1.4 were designed to verify the influence of pore volume along with the possible effects of decreasing the distance between pore boundary and sample edge. Samples 1.3 and 1.5 contain one prolate ellipsoid and one sphere, respectively. The sizes of voids are designed in such a way as to overcome the limitation of printer resolution and make the softening significant enough for it to be easily measurable.  
The samples were printed using FDM (Prusa i3 MK3S+ printer) and SLA (Form3+ printer) technologies. In the case of FDM printing, we used PLA as the filament material. We set 100\% rectilinear infill and select two different options for wall thicknesses (perimeters); our samples have either two or five perimeters. 
In Figure~\ref{samples_starters_FDM}, we present half-printed samples having two perimeters. It can be seen that the low aspect ratio of cracks in samples 1.2 and 1.4 leads to some printing imperfections.
Also, the visual effect that the number of perimeters has on the edges of both the sample and the internal void can be appreciated in Figure~\ref{perimeters}.
\par
\begin{figure}[!htbp]
\centering
\includegraphics[scale=0.215]{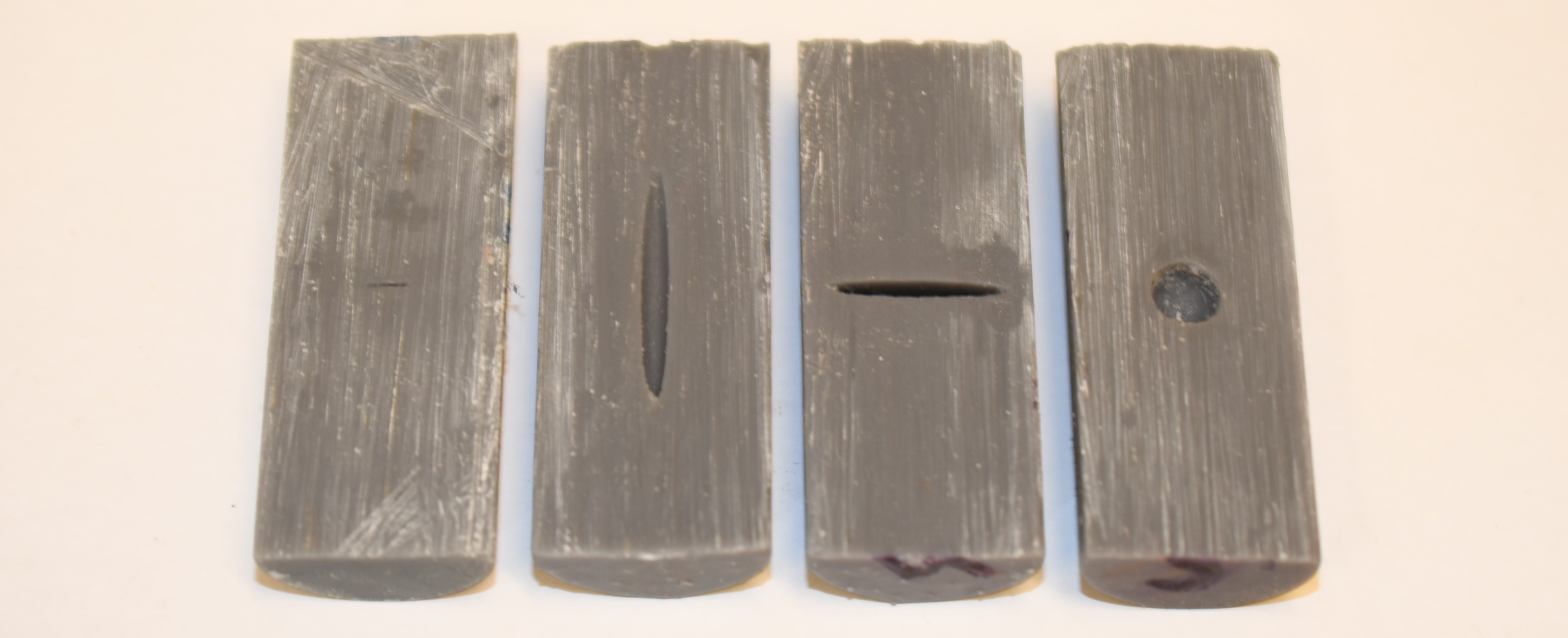}
\caption{\footnotesize{Printed and deformed SLA samples cut in half with voids cleaned} }
\label{samples_starters_SLA}
\end{figure}
\begin{figure}[!htbp]
\centering
\includegraphics[scale=0.2]{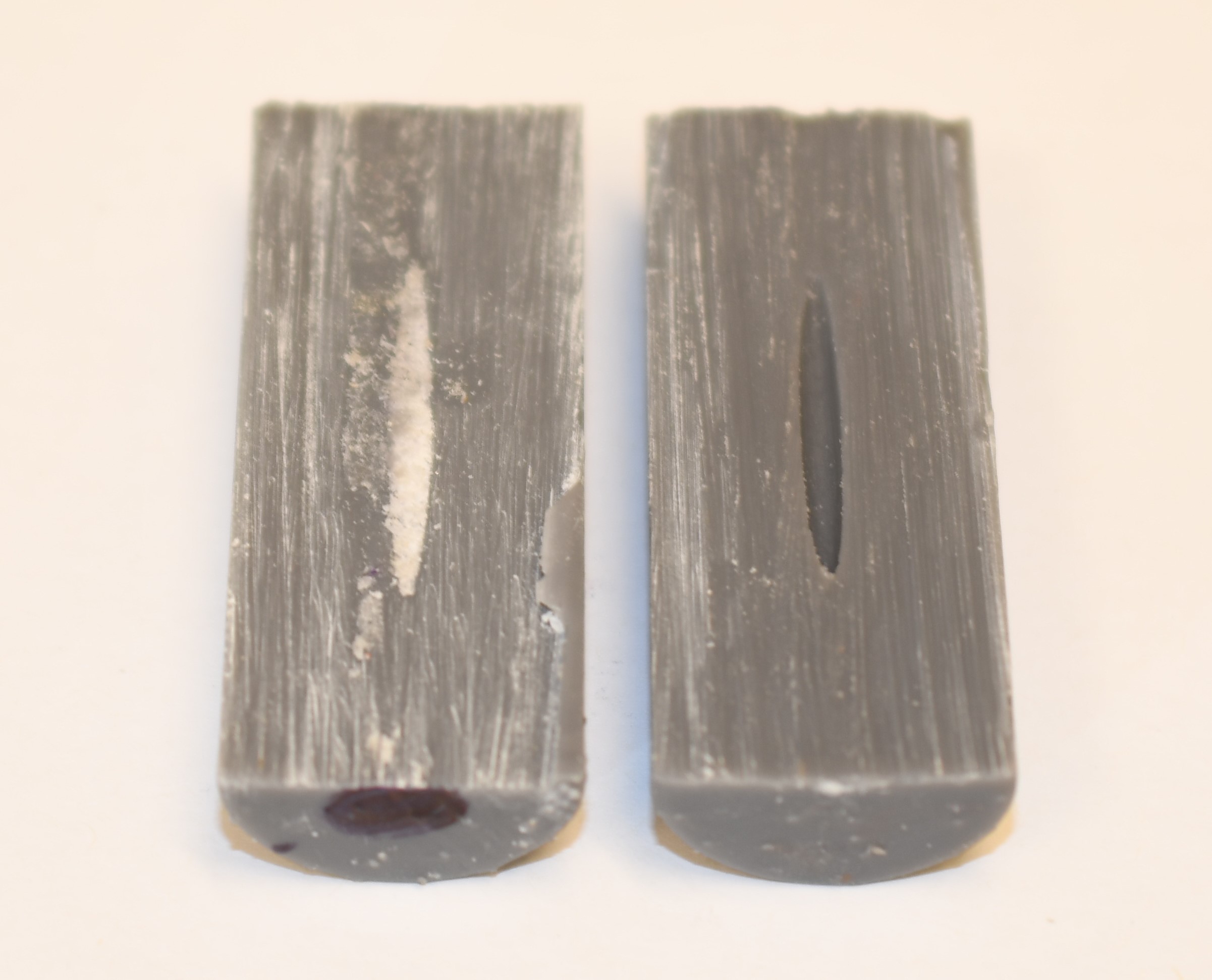}
\caption{\footnotesize{Residual resin present in the uncleaned void of SLA.}}
\label{residual_resin}
\end{figure}

In the case of the SLA printing, we used standard grey resin.
The manufacturer suggests putting freshly printed material in IPA (alcohol) and heating it to make the fabric stiffer. In this paper, we go beyond this suggestion and check two other procedures: a) IPA cleaning and cure for a month without heating, b) IPA cleaning and cure for a week without heating, and c) IPA cleaning, heating ($60^{\circ}$C), and fast cure (between 4 hours to 2 days). This way, we verify the influence of a cure time on the sample stiffness and compare it with the influence of the heating.  
In Figure~\ref{samples_starters_SLA}, we present cleaned SLA samples cut in half. While cutting them, we noticed the remaining residual resin inside the voids that may have an influence on sample stiffness, as shown in Figure~\ref{residual_resin}.
\begin{table}[!htbp]
\scalebox{1}{
\begin{tabular}
{cccccc}
\toprule
sample & void type & radius [mm] & aspect ratio & number of voids  & orientation \\
\toprule
1.1& --- & --- & --- & 0 & ---  \\
\cmidrule{1-6}
1.2& oblate spheroid & 2.5 & 0.1 & 1 & horizontal    \\
\cmidrule{1-6}
1.3& prolate spheroid & 1.5 & 10 & 1 & vertical   \\
\cmidrule{1-6}
1.4& oblate spheroid & 10 & 0.1 & 1 & horizontal  \\
\cmidrule{1-6}
1.5& sphere & 4 & 1 & 1 & --- \\
\bottomrule
\end{tabular}
}
\caption{\footnotesize{Description of samples designed for the preliminary phase of the experiments. All voids are placed in the centre of the sample. }}
\label{tab:phase_one}
\end{table}


\begin{figure}[!htbp]
\centering
\includegraphics[scale=0.5]{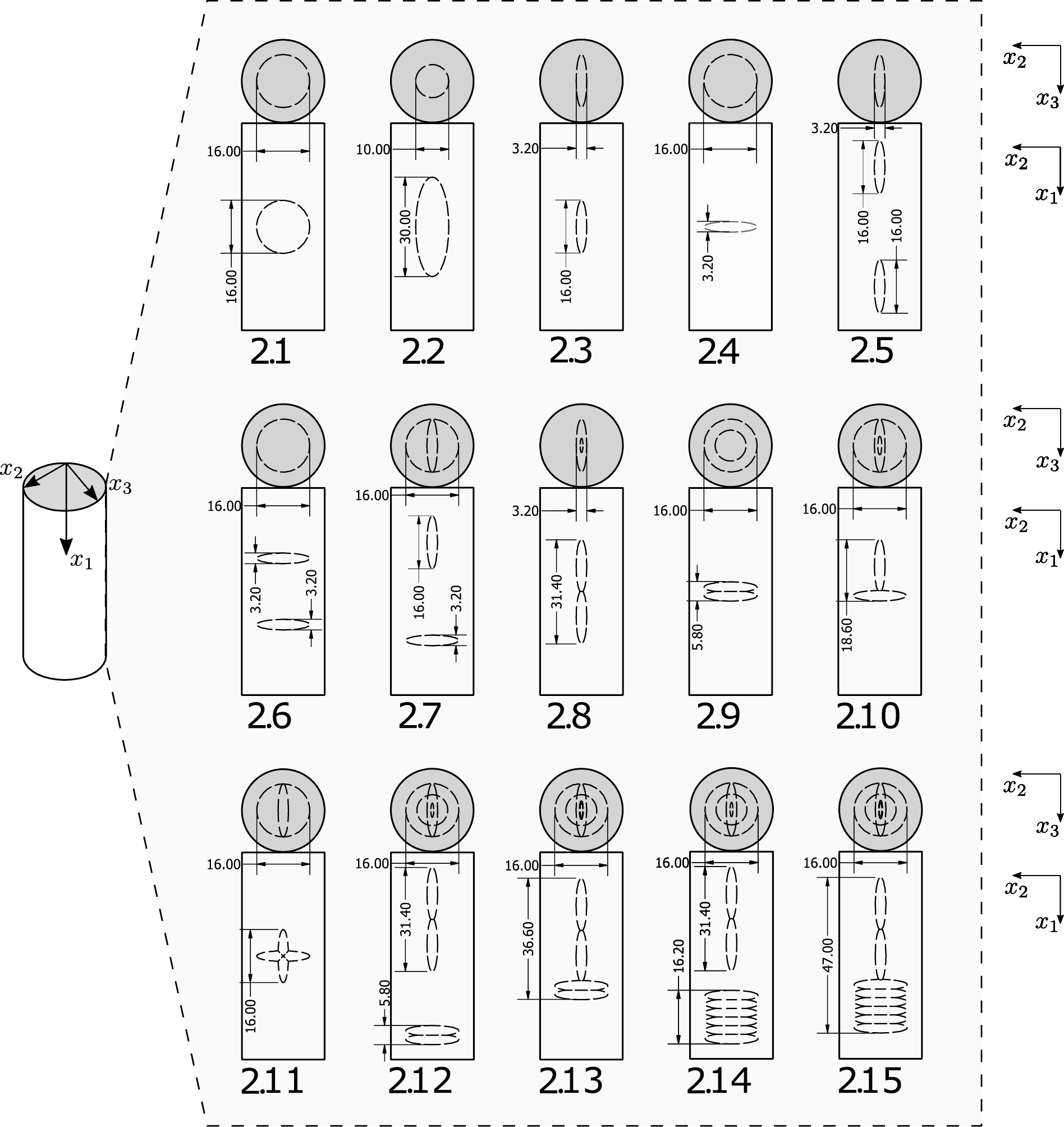}
\caption{\footnotesize{Geometry of samples designed for the final phase of experiments, where dimensions are in millimetres. Samples 2.3--2.15 consist of either isolated, overlapping, or crossing oblate spheroids having a $0.2$ aspect ratio and $8$mm of radius. }}
\label{fig:phase_two}
\end{figure}

\begin{figure}[!htbp]
\centering
\begin{subfigure}{.3\textwidth}
\includegraphics[scale=0.14]{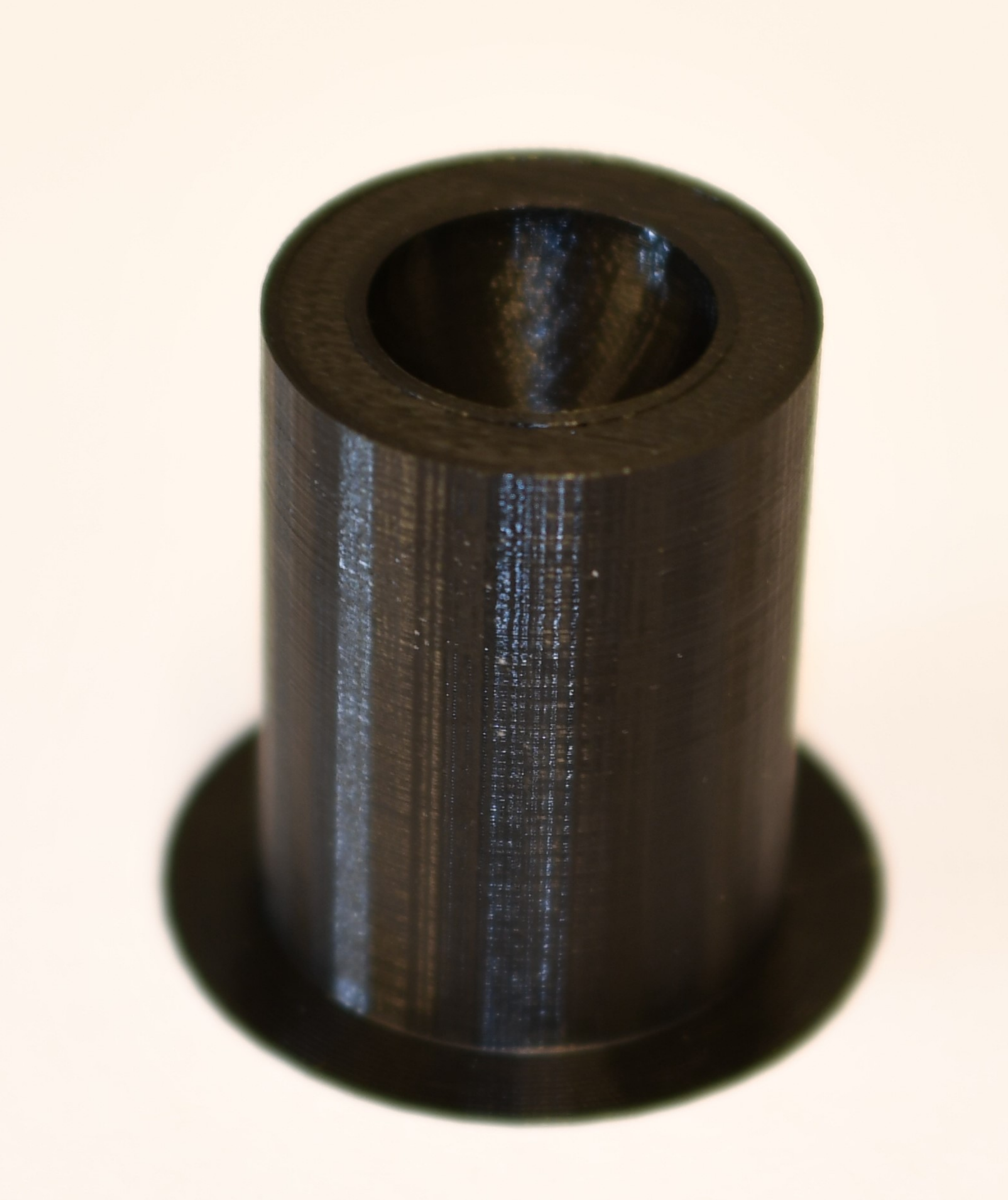}
\caption{}
\end{subfigure}
\begin{subfigure}{.3\textwidth}
\includegraphics[scale=0.14]{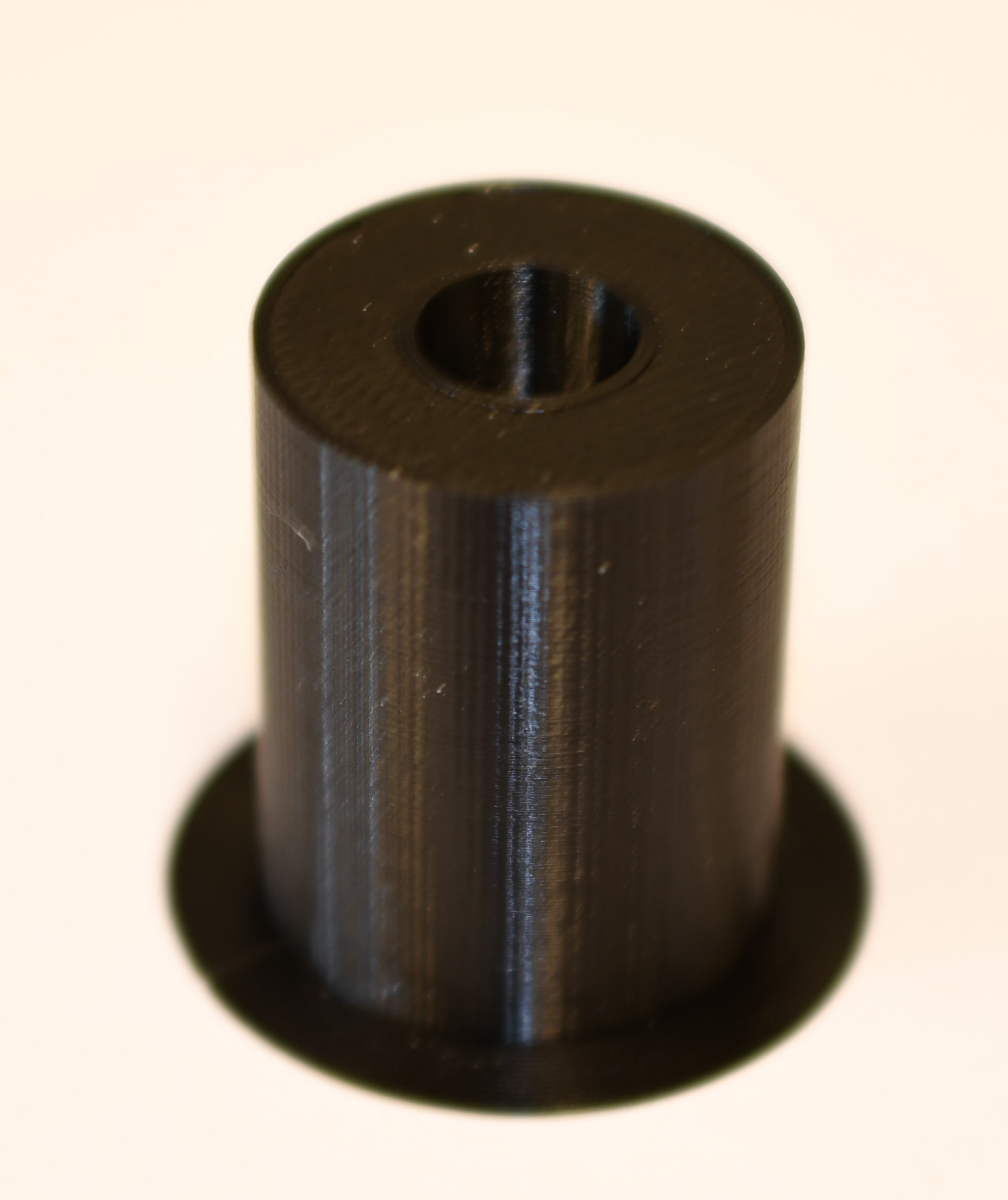}
\caption{}
\end{subfigure}
\begin{subfigure}{.3\textwidth}
\includegraphics[scale=0.14]{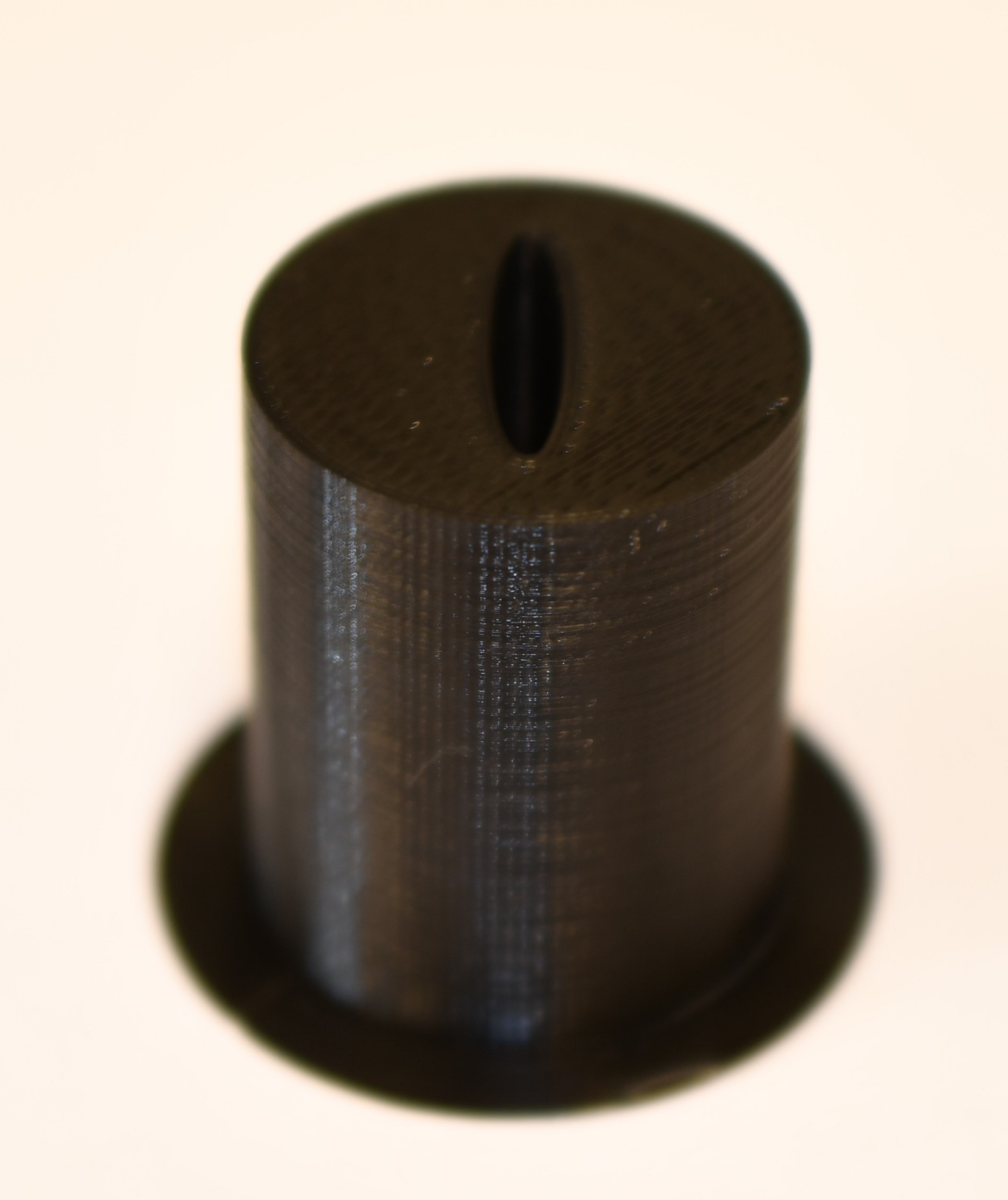}
\caption{}
\end{subfigure}
\begin{subfigure}{.3\textwidth}
\includegraphics[scale=0.14]{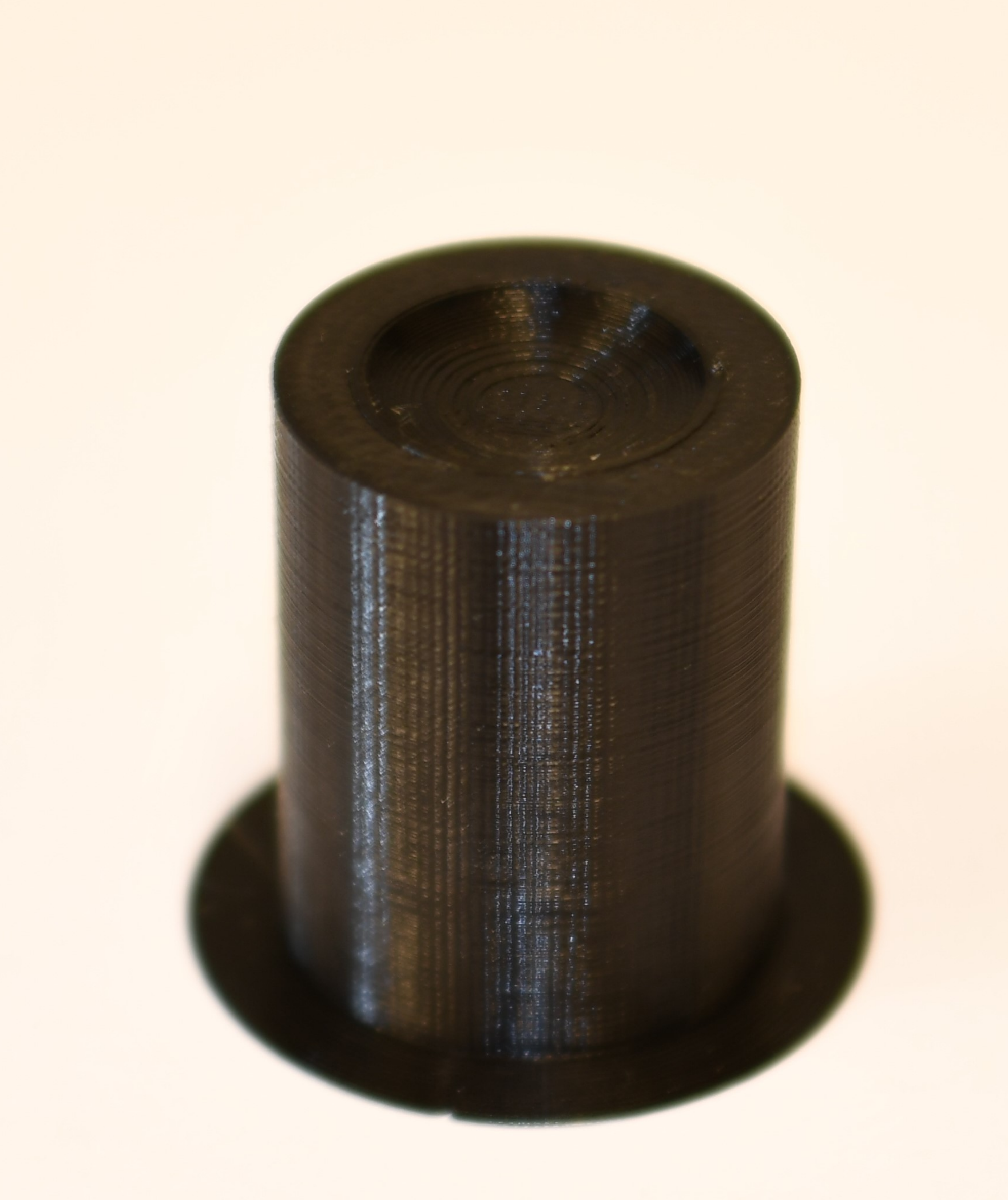}
\caption{}
\end{subfigure}
\begin{subfigure}{.3\textwidth}
\includegraphics[scale=0.14]{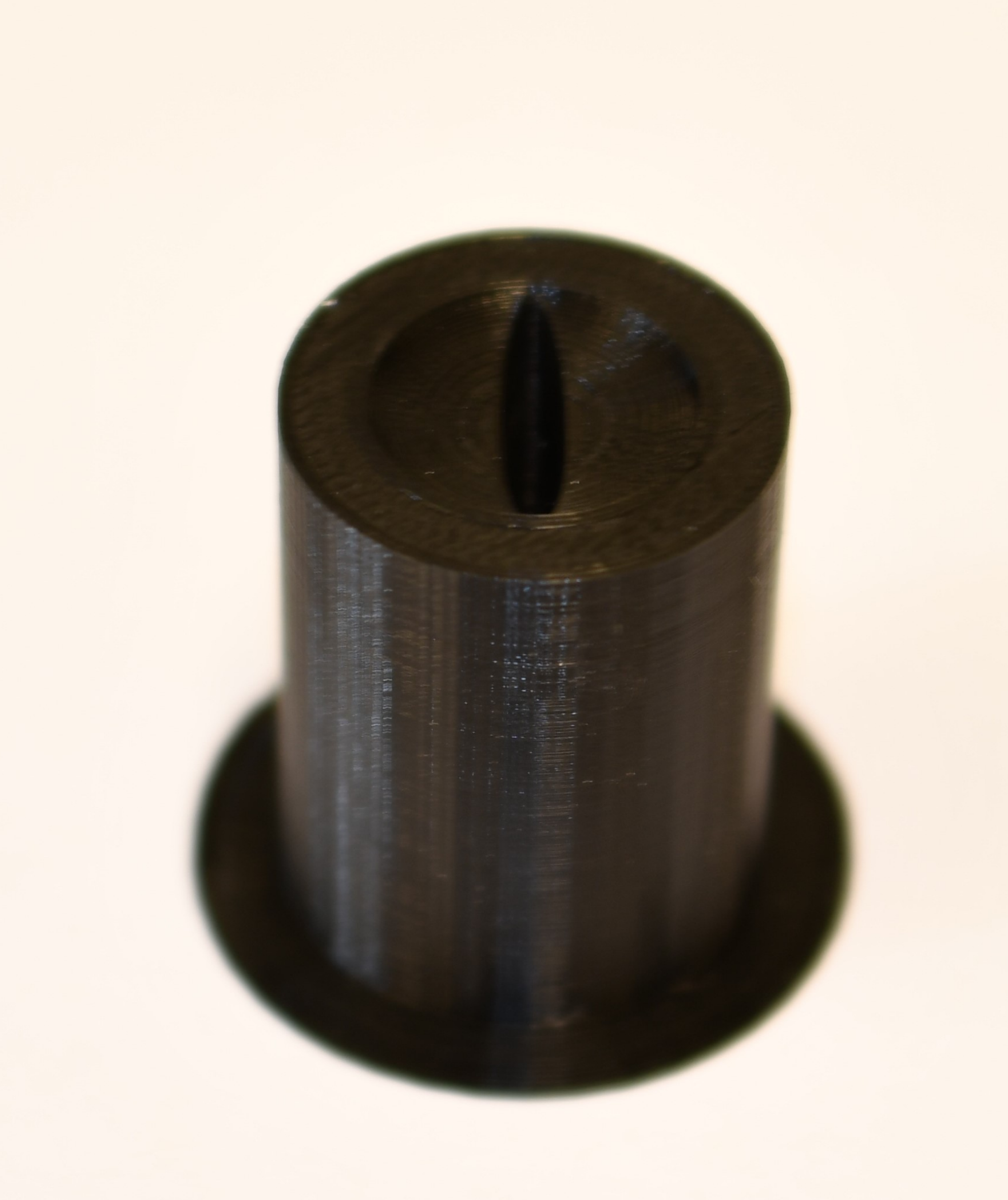}
\caption{}
\end{subfigure}
\begin{subfigure}{.3\textwidth}
\includegraphics[scale=0.14]{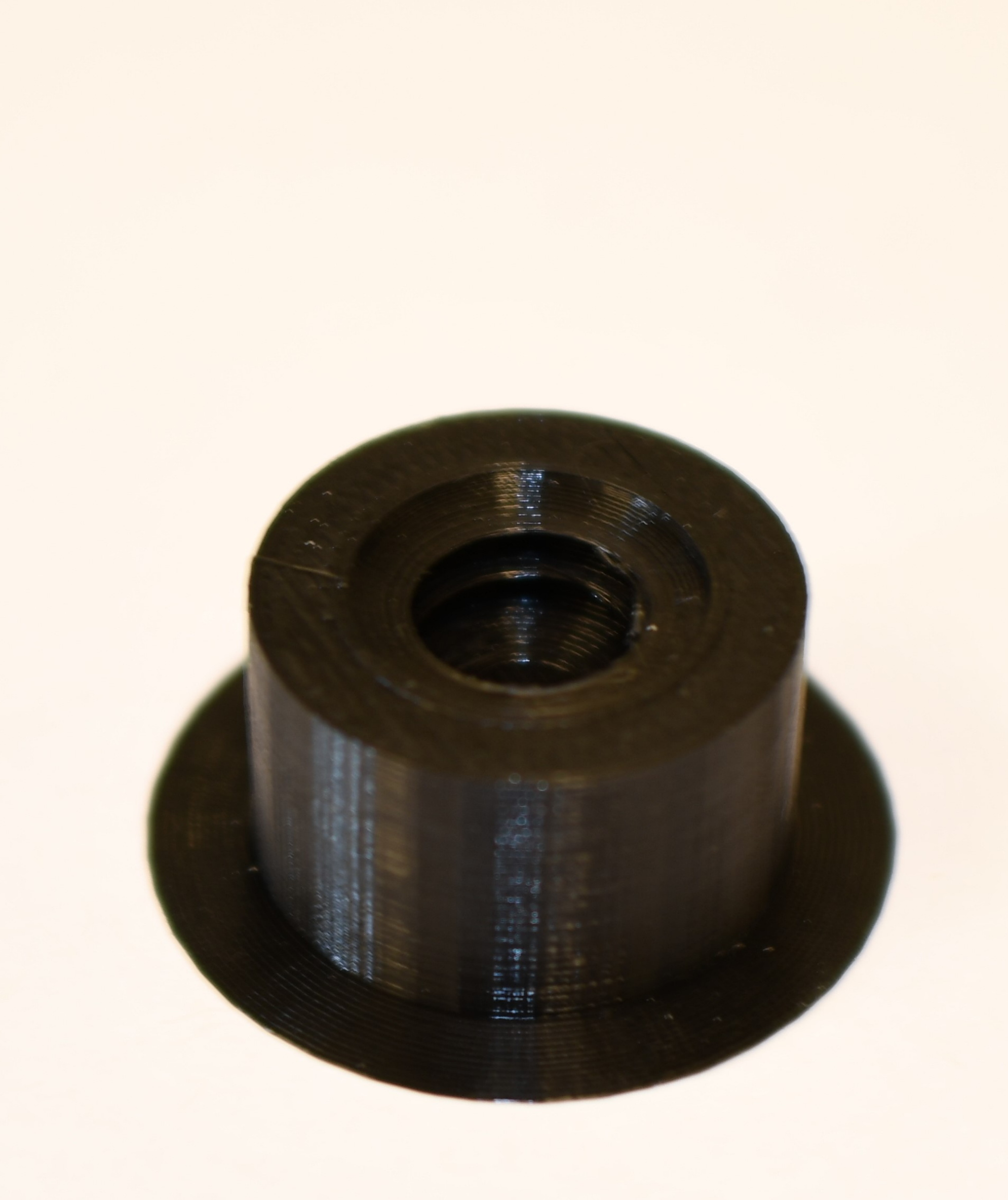}
\caption{}
\end{subfigure}
\caption{\footnotesize{Half-printed FDM samples having two perimeters, used in the final phase of experiments. Each type of printed void is shown. These configurations are present in {\bf{a)}} sample 2.1, {\bf{b)}} sample 2.2, {\bf{c)}} samples 2.3, 2.5, 2.7, 2.8, 2.10, 2.12--2.15, {\bf{d)}} samples 2.4, 2.6, 2.7, 2.9, 2.10, 2.12--2.15, {\bf{e)}} sample 2.11, and {\bf{f)}} samples 2.14--2.15. All cores have brims that prevent their detachment from the bed.}}
\label{precious_samples}
\end{figure}

The longitudinal and cross-sections of samples designed for the final phase of experiments are depicted in Figure~\ref{fig:phase_two}; the concise description of voids' geometries can be found in Table~\ref{tab:phase_two}. Samples 2.1 and 2.2 contain a sphere and a prolate spheroid, respectively. They are similar to samples 1.5 and 1.3, but the volume fraction of the voids is larger to augment the softening effect. In the remaining samples, 2.3--2.15, we focus on crack-like shapes that are crucial in the context of e.g., geomaterials. 
These samples consist of either isolated, overlapping, or crossing oblate spheroids having a $0.2$ aspect ratio and radius of $8$ mm. 
The radius of the voids is smaller than that of sample 1.4; this way, the edge effect is diminished and potential collapse of the crack is inhibited. By contrast, the aspect ratio is augmented to a) avoid printing imperfections, b) keep a significant softening effect, c) keep it flat enough to be considered a crack, and d) allow fluid injection in the future. Samples 2.3 and 2.4  are designed to verify the influence of a crack-like void orientation. We designed other samples to verify the possible influence of interactions between two vertical cracks (samples 2.5 and 2.8), horizontal cracks (samples 2.6 and 2.9), and vertical and horizontal cracks (samples 2.7, 2.10, and 2.11). Further, samples 2.12--2.15 may allow us to measure the significant softening effect (as compared to plain sample 1.1) and interaction influence of a moderate crack concentration; the total volume fraction of voids in samples 2.12--2.13 is $\phi\approx5.6\%$ and in samples 2.14--2.15 is $\phi\approx11.2\%$. Note that the overlapping of the samples is small enough to keep the crack-like shapes but perhaps sufficient to equilibrate pore pressure in future fluid injection experiments.
In this phase of experiments, all samples are printed using FDM technology and have two perimeters. The interior of each void (either isolated, overlapping, or crossing) type is shown in Figure~\ref{precious_samples}.
\begin{table}[!htbp]
\scalebox{1}{
\begin{tabular}
{ccccc}
\toprule
sample & void type(s) & number of voids  & void centre to sample centre [mm] & orientation(s) \\
\toprule
2.1& sphere, $a=8$mm  & 1& 0 & ---   \\
\cmidrule{1-5}
\multirow{2}{*}{2.2}& prolate spheroid & \multirow{2}{*}{1} &\multirow{2}{*}{0}&  \multirow{2}{*}{vrt.} \\
& $a=5$mm, $\gamma=3$ &  & &   \\
\cmidrule{1-5}
2.3& oblate spheroid & 1  & 0 & vert.   \\
\cmidrule{1-5}
2.4& oblate spheroid & 1  & 0 & hor.   \\
\cmidrule{1-5}
2.5& oblate spheroids & 2  & 18/-18 & vrt./vrt.   \\
\cmidrule{1-5}
2.6& oblate spheroids & 2  & 11.6/-11.6 & hor./hor.   \\
\cmidrule{1-5}
2.7& oblate spheroids & 2  & 14.8/-14.8 & vrt./hor.   \\
\cmidrule{1-5}
2.8& oblate spheroids & 2  & 7.7/-7.7 & vrt./vrt.   \\
\cmidrule{1-5}
2.9& oblate spheroids & 2  & 1.3/-1.3 & hor./hor.   \\
\cmidrule{1-5}
2.10& oblate spheroids & 2  & 7.7/-1.3 & vrt./hor.   \\
\cmidrule{1-5}
2.11& oblate spheroids & 2  & 0/0 & vrt./hor.   \\
\cmidrule{1-5}
2.12  & oblate spheroids & 4 & 18.75/3.35/-22.55/-25.15 & vrt./vrt./hor./hor.   \\
\cmidrule{1-5}
2.13  & oblate spheroids & 4 & 15.4/0/-9/-11.6 & vrt./vrt./hor./hor.   \\
\cmidrule{1-5}
\multirow{2}{*}{2.14}  & \multirow{2}{*}{oblate spheroids} & \multirow{2}{*}{8}  &18.75/3.35/-12.15/-14.75/  &  \multirow{2}{*}{vrt.x2/hor.x6} \\
 & &  &-17.35/-19.95/-22.55/-25.15  &   \\
\cmidrule{1-5}
\multirow{2}{*}{2.15}  & \multirow{2}{*}{oblate spheroids} & \multirow{2}{*}{8}  &15.5/0.1/-8.9/-11.5/  &  \multirow{2}{*}{vrt.x2/hor.x6} \\
 & &  &-14.1/-16.7/-19.3/-21.9  &   \\
\bottomrule
\end{tabular}
}
\caption{\footnotesize{Description of samples designed for the final phase of experiments. All voids are centred horizontally. The distance measured from the void centre towards the bottom of the sample is positive. Orientations of the voids are listed in descending order starting from the uppermost void. All oblate spheroids have a radius $a=8$mm and aspect ratio $\gamma=0.2$. }}
\label{tab:phase_two}
\end{table}
%
\newpage
\subsection{Moduli definitions and measurement repetitions}\label{sec:mod}
As mentioned in Section~\ref{sec:emt}, we measure either Young's modulus and Poisson's ratio (plain samples) or Young's modulus only (samples with voids).
We subjectively choose four pairs of start and end points on the strain-stress curve to define the limits for Young's moduli:
\begin{itemize}
\item{15\% and 30\% of the yield strength, $E_{\rm{15-30}}$,}
\item{25\% and 50\% of the yield strength, $E_{\rm{25-50}}$,}
\item{25\% and 75\% of the yield strength, $E_{\rm{25-75}}$,}
\item{lower and upper elastic limits, $E_{\rm{el}}$.}
\end{itemize}
For instance,
\begin{equation}
E_{{15}-{30}}=\frac{0.30\,\sigma_{y}-0.15\,\sigma_{y}}{u(0.30\,\sigma_{y})-u(0.15\,\sigma_{y})}\times l,
\end{equation}
where $\sigma_y$ is the stress at yield, $u$ is the displacement measured by LVDT, $l$ is the sample's average length.
In the case of FDM, the yield strength is determined based on the first peak of the stress-strain curve. 
In the case of SLA, yield is difficult to determine unequivocally due to the continuous growth of stress with displacement. 
Therefore, we arbitrarily choose a threshold of 6\% strain to obtain the reference point (yield strength approximation) used to calculate the limits of Young's modulus. 
To define the Poisson's ratio, we first estimate two ranges of stresses that correspond to the elastic part of axial and radial curves. Second, we choose a common range of stress that---in a set theory sense---is the intersection of the two aforementioned ranges. This way, the parts of $\varepsilon_1$ and $\varepsilon_r$ needed for Poisson's ratio computation are unequivocally determined. 

Our compression experiments and modulus measurements are less prone to random errors due to the repetitions of 3D printing and testing of identically designed samples. This way, more reliable average values of moduli can be obtained. Table~\ref{tab:repetitions} lists the number of repetitions of valid measurements for each sample type of this study. 
Note that the plain sample (1.1) has the most repetitions; this is due to its importance for the EMT calculations. 
\begin{table}[!htbp]
\scalebox{1}{
\begin{tabular}
{lcccc}
 & \multicolumn{3}{c}{YM valid measurements} & PR valid measurements\\
\cmidrule{2-5}
 & sample 1.1 & sample 1.2--1.5 & sample 2.1--2.15  & sample 1.1\\
\bottomrule 
& & & &     \\ [-0.13in]  
${\rm{FDM}}_{\rm{t}}$ (two perim.)& 12 & 6 & 6  & 5 \\  
\cmidrule{1-5}
${\rm{FDM}}_{\rm{v}}$ (five perim.)& 12 & 6 & ---  & 4 \\
\cmidrule{1-5}
${\rm{SLA}}_{\rm{m}}$ (month cure)& 3 & 3 & ---  & 3 \\
\cmidrule{1-5}
${\rm{SLA}}_{\rm{w}}$ (week cure) & 3 & 3 & ---  &  3 \\
\cmidrule{1-5}
${\rm{SLA}}_{\rm{f}}$ (fast cure, heated)& 3 & 3 & ---  & 3 \\
\bottomrule
\end{tabular}
}
\caption{\footnotesize{Number of repetitions of valid Young's moduli and Poisson's ratio measurements for each sample type }}
\label{tab:repetitions}
\end{table}

%

\section{Results}
\subsection{Stress-strain curves}\label{stressstrainsec}

\begin{figure}[!htbp]
\centering
\begin{subfigure}{.45\textwidth}
\includegraphics[scale=0.6]{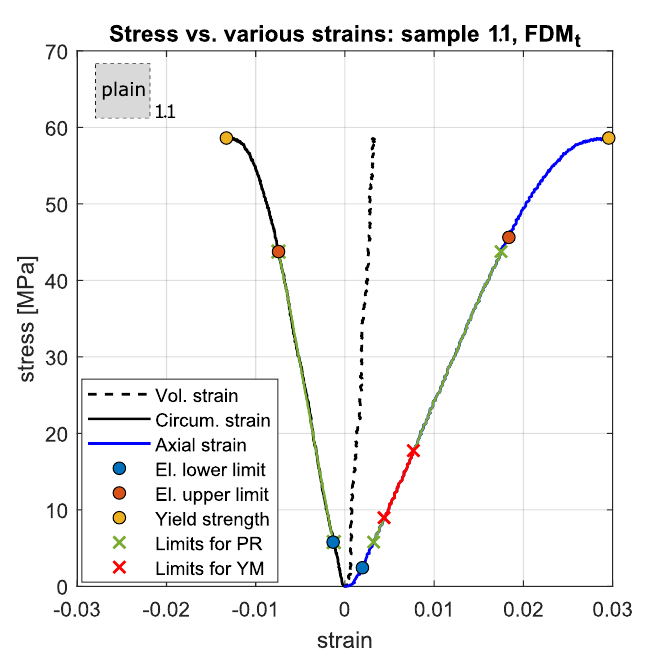}
\end{subfigure}
\qquad
\begin{subfigure}{.45\textwidth}
\includegraphics[scale=0.6]{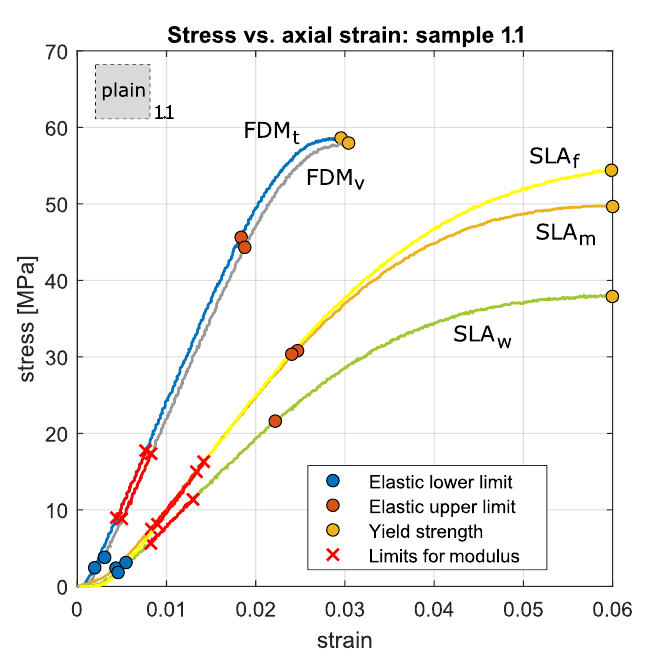}
\end{subfigure}
\begin{subfigure}{.45\textwidth}
\includegraphics[scale=0.6]{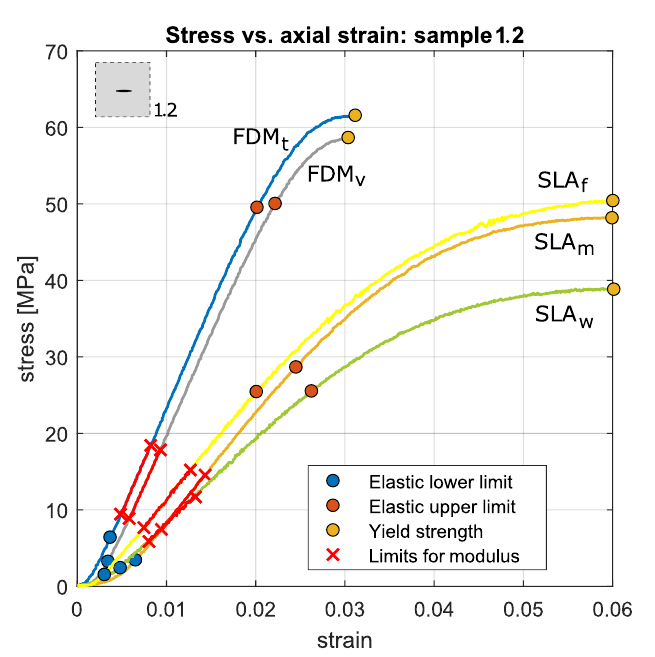}
\end{subfigure}
\qquad
\begin{subfigure}{.45\textwidth}
\includegraphics[scale=0.6]{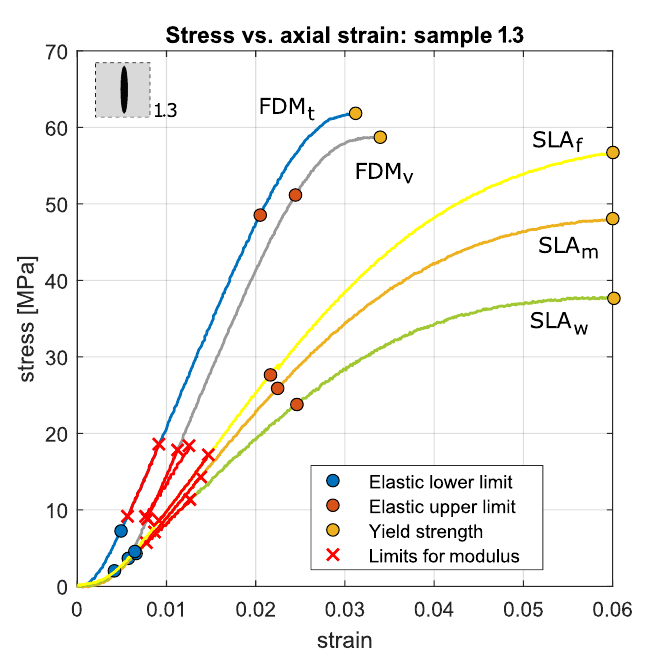}
\end{subfigure}
\begin{subfigure}{.45\textwidth}
\includegraphics[scale=0.6]{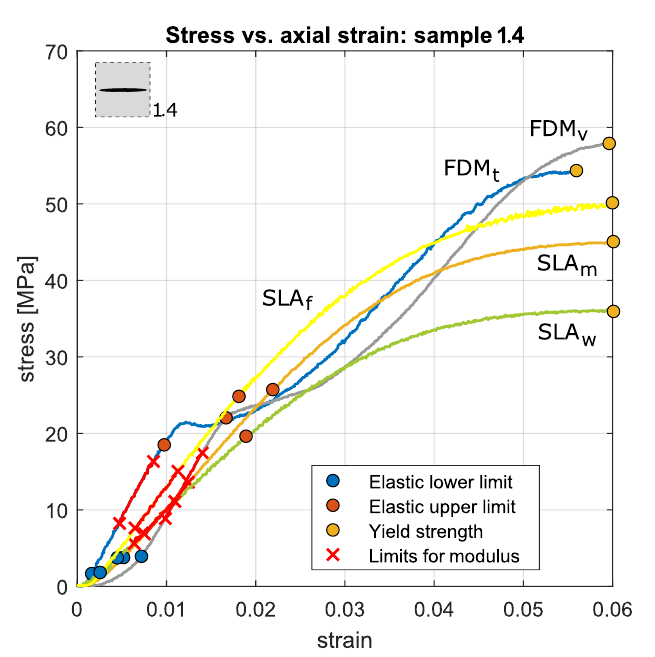}
\end{subfigure}
\qquad
\begin{subfigure}{.45\textwidth}
\includegraphics[scale=0.6]{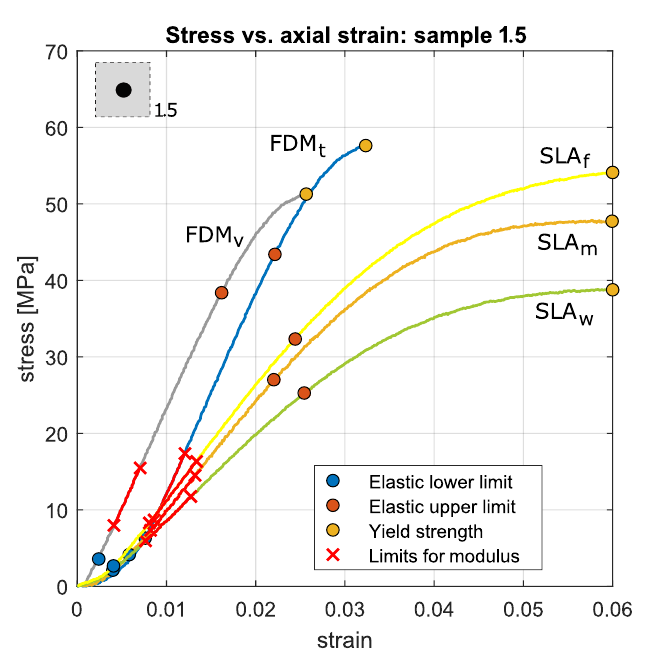}
\end{subfigure}
\caption{\footnotesize{Selected stress-strain curves for samples 1.1-1.5 that are FDM with two perimeters (${\rm{FDM}}_{\rm{t}}$) or five perimeters (${\rm{FDM}}_{\rm{v}}$), SLA cured for a month (${\rm{SLA}}_{\rm{m}}$), a week (${\rm{SLA}}_{\rm{w}}$), or heated and fast-cured for less than two days (${\rm{SLA}}_{\rm{f}}$). Elastic parts, yield points and the range of curve to obtain $E_{15-30}$ are presented. Additionally, the left-upper figure presents the circumferential and volumetric strains of plain FDM sample having two perimeters (${\rm{FDM}}_{\rm{t}}$). The Poisson's ratio is obtained based on the elastic portion of the circumferential strain that lies within elastic part of the axial strain.
}}
\label{stresstrain}
\end{figure}

Results of axial strain as a function of stress are plotted for the five preliminary-phase microstructural configurations (samples 1.1-1.5) (Figure \ref{stresstrain}). Selected curves are plotted for a) FDM printed samples with 2 perimeters (${\rm{FDM}}_{\rm{t}}$), b) FDM printed samples with 5 perimeters (${\rm{FDM}}_{\rm{v}}$, c) SLA printed samples tested after month of post-printing curing time at room temperature (${\rm{SLA}}_{\rm{m}}$), d) SLA after a week of curing time at ambient temperature (${\rm{SLA}}_{\rm{w}}$) and e) SLA after curing in a heated chamber at $60^{\circ}$C for one hour, and then post-curing at ambient temperature for a minimum time of four hours and a maximum of two days (${\rm{SLA}}_{\rm{f}}$). Curves are plotted up to their yield points, displaying their elastic ranges and the portions of the curves considered for elasticity modulus computation.  Circumferential strain as a function of stress is plotted for a plain ${\rm{FDM}}_{\rm{t}}$ sample, displaying the ranges considered for Poisson's ratio determination. 

It can be seen that FDM printed samples are stiffer and stonger with respect to SLA samples, and that the printing pattern (numbers of delimiting perimeters) does not significantly affect their mechanical behavior, although samples printed with two perimeters ($\rm{FDM}_{\rm{t}}$) consistently display higher stiffness and higher yield strength with respect to those with five delimiting contours (${\rm{FDM}}_{\rm{v}}$) for all preliminary-phase samples except for 1.4.
SLA printed samples are relatively soft, elasto-plastic materials. Curing time and method have a significant influence on the mechanical properties of SLA printed samples. Two days of curing time within a heated chamber at $60^{\circ}$C is shown to be the most effective method to drive the material to higher mechanical strength. In addition, samples cured for one month are consistently stiffer and stronger than samples cured for one week. 

\begin{figure}[!htbp]
\centering
\includegraphics[scale=0.12]{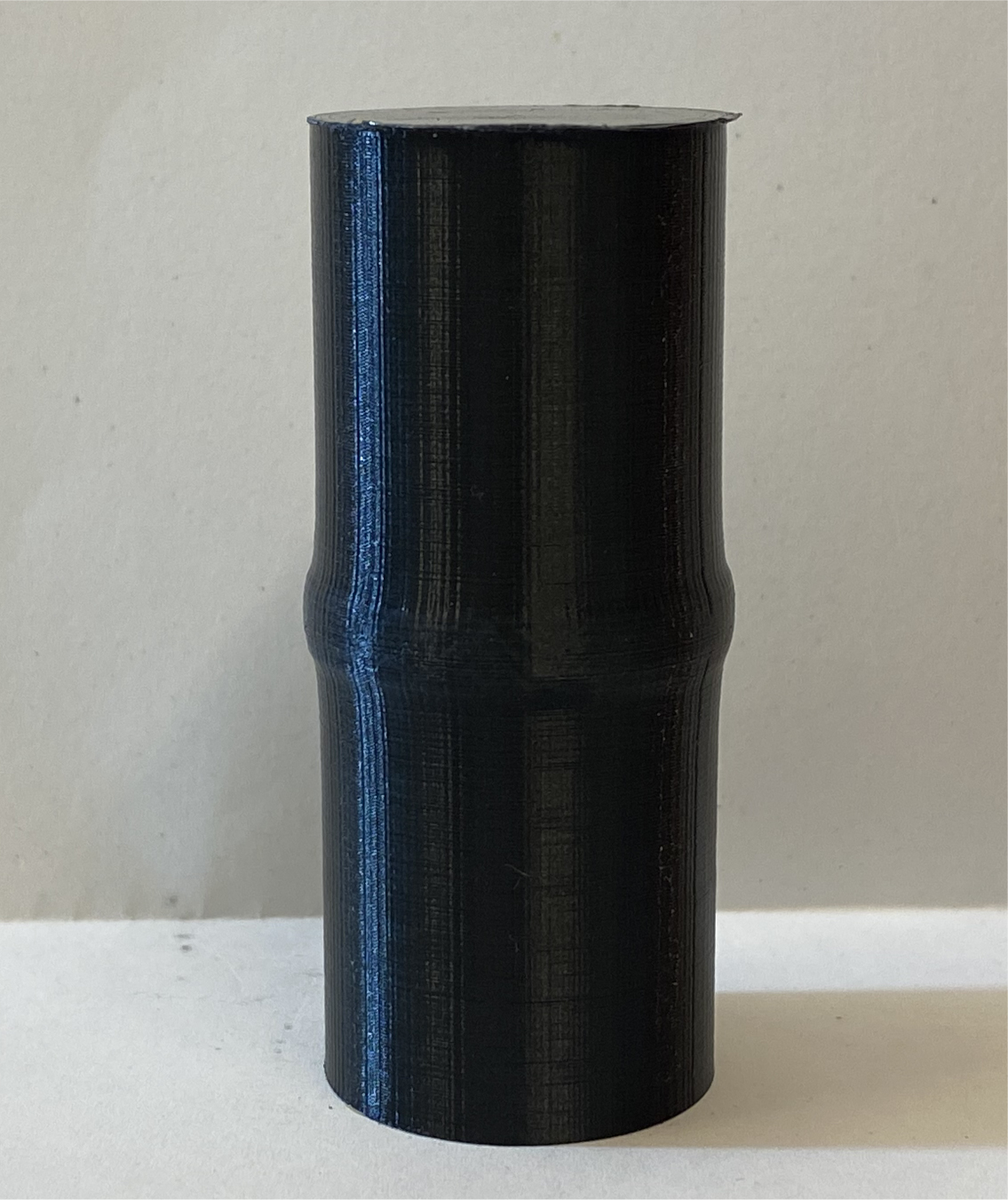}
\label{sample03FDM}
\caption{\footnotesize{Inelastic deformation mechanism due to pore space collapse in FDM printed sample with a large oblate pore (1.4). }}
\label{fig:edge_effect2}
\end{figure}

The microstructural configuration of samples does not significantly affect the mechanical properties of preliminary-phase 3D printed samples, except for the 1.4 microstructural design, which notably alters the deformation mechanism of FDM printed samples. This design has the largest void fraction and smallest distance from void to sample edge. Predictably, this microstructural configuration induces a significant softening effect and causes a different deformation mechanism by which the pore boundaries collapse at lower stress due to stress concentrations at the thin walls surrounding the pore (Figure ~\ref{fig:edge_effect2}).

Poisson's ratio measurements for plain FDM samples printed with two and five perimeters (${\rm{FDM}}_{\rm{t}}$, ${\rm{FDM}}_{\rm{v}}$) and SLA samples with the previously described curing methods (${\rm{SLA}}_{\rm{m}}$, ${\rm{SLA}}_{\rm{w}}$, ${\rm{SLA}}_{\rm{f}}$)  are indicated in Table \ref{tab:poissons}.

\begin{table}[!htbp]
\scalebox{1}{
\begin{tabular}
{llllc}
 ${\rm{FDM}}_{\rm{t}}$ & ${\rm{FDM}}_{\rm{v}}$ & ${\rm{SLA}}_{\rm{m}}$ & ${\rm{SLA}}_{\rm{w}}$ & ${\rm{SLA}}_{\rm{f}}$\\
\toprule
$\nu_0=0.41$ & $\nu_0=0.38$ & $\nu_0=0.42$ & $\nu_0=0.42$ & $\nu_0=0.45$\\

\bottomrule
\end{tabular}
}
\caption{\footnotesize{Poisson's ratio for plain FDM and SLA samples.}}
\label{tab:poissons}
\end{table}

\subsection{Mean Young's modulus of plain FDM samples} \label{meanmodulus}

\begin{figure}[!htbp]
\centering
\begin{subfigure}{.45\textwidth}
\includegraphics[scale=0.7]{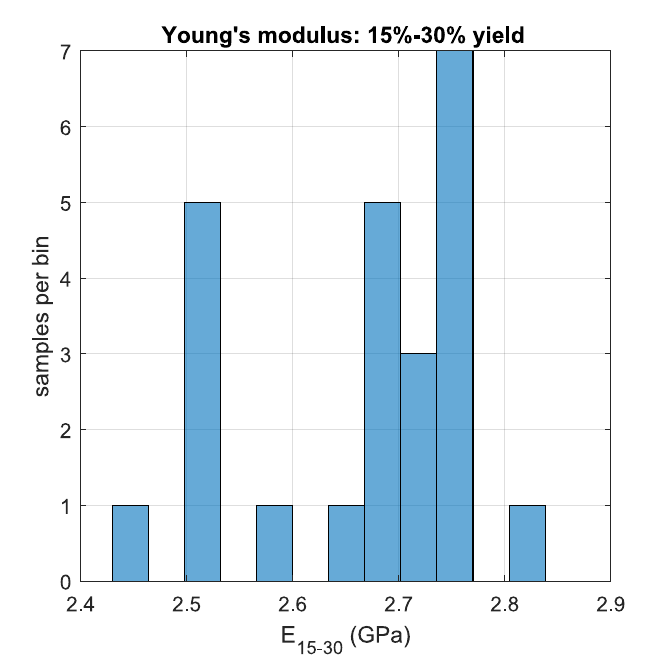}
\end{subfigure}
\qquad
\begin{subfigure}{.45\textwidth}
\includegraphics[scale=0.7]{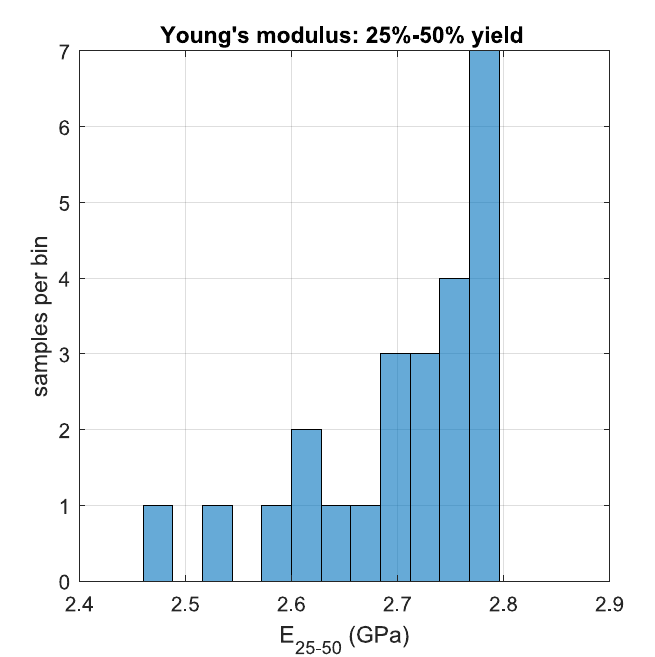}
\end{subfigure}
\begin{subfigure}{.45\textwidth}
\includegraphics[scale=0.7]{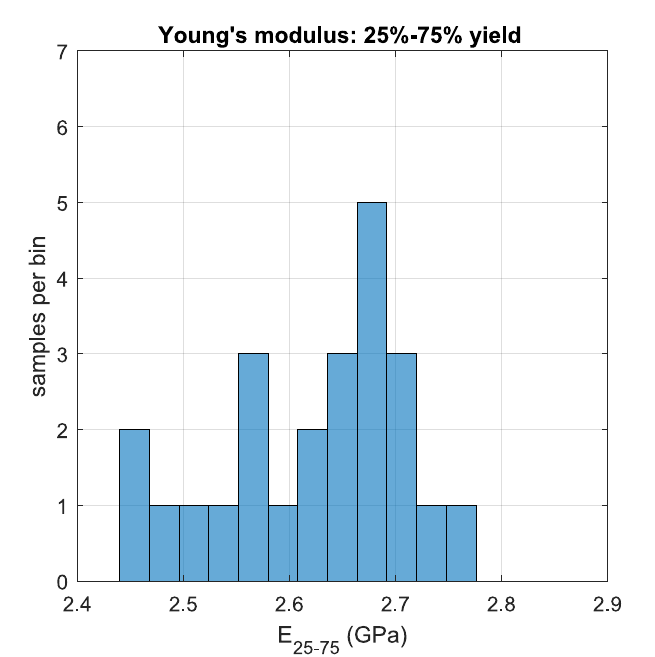}
\end{subfigure}
\qquad
\begin{subfigure}{.45\textwidth}
\includegraphics[scale=0.7]{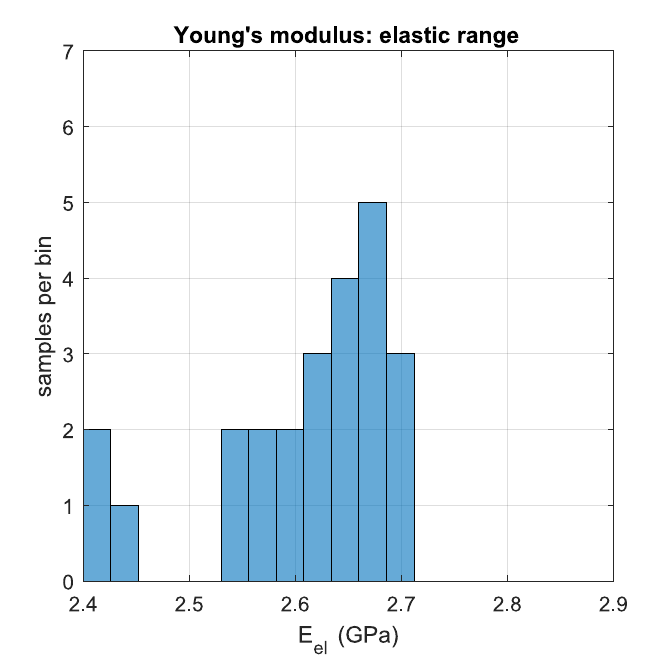}
\end{subfigure}
\caption{\footnotesize{Distributions of experimentally obtained Young's modulus for plain FDM samples having either two or five perimeters. In total, there were twenty--four samples tested. Results in each chart are grouped into twelve bins.}}
\label{fdmaveragemodulus}
\end{figure}

The average Young's modulus calculated from the arithmetic mean of 12 and 6 repeated tests for plain FDM samples, and samples 1.2---1.5, respectively, are shown in Figure \ref{fdmaveragemodulus} and listed in Table \ref{tab:distr} for the four modulus determination methods described in section \ref{moduli}. Young's modulus of plain samples consistently lie between $2.6$ and $2.7 \rm{GPa}$, and display a scatter of $\pm 0.2 \rm{GPa}$. Results do not follow a Gaussian distribution, although this is likely due to insufficient number of tested samples. These results highlight the need for considering means calculated from a significant statistical sample size to obtain meaningful and representative results for the elasticity modulus of 3D printed samples.   

Young's modulus results vary depending on the portion of the elastic range chosen for modulus computation. The higher magnitudes are obtained using the $25$ to $50\%$ portion of the elastic range. ${\rm{FDM}}_{\rm{t}}$ samples display a consistently higher Young's modulus with respect to ${\rm{FDM}}_{\rm{v}}$, although the difference is not significant ($0.05 \rm{GPa}$ or less).

\begin{table}[!htbp]
\scalebox{1}{
\begin{tabular}
{lcccc}
\toprule
\multicolumn{2}{c}{} & ${1.1}_t$  & ${1.1}_v$ & ${1.1}_t$ \& ${1.1}_v$ \\
\toprule
$E_{15-30}$ & mean & 2.69 & 2.64 & 2.67  \\
& range & $2.52-2.83$ & $2.44-2.77$  & $2.44-2.83$   \\
\cmidrule{1-5}
$E_{25-50}$& mean & 2.73 & 2.68 & 2.71  \\
& range & $2.61-2.79$ & $2.46-2.77$  & $2.46-2.79$   \\
\cmidrule{1-5}
$E_{25-75}$& mean & 2.64 & 2.61 & 2.62  \\
& range & $2.50-2.73$ & $2.44-2.77$  & $2.44-2.77$   \\
\cmidrule{1-5}
$E_{\rm{el}}$& mean & 2.63 & 2.59 & 2.61  \\
& range & $2.54-2.70$ & $2.41-2.70$  & $2.41-2.70$   \\
\toprule
\end{tabular}
}
\caption{\footnotesize{Experimentally obtained Young's modulus for plain FDM samples having either two (sample ${1.1}_t$, twelve tests) or five perimeters (sample ${1.1}_v$, twelve tests). Mean values and ranges of differently defined modulus are given.}}
\label{tab:distr}
\end{table}
\subsection{Effect of post-printing curing methods for the mechanical behaviour of SLA samples}

\begin{figure}[!htbp]
\centering
\begin{subfigure}{.45\textwidth}
\includegraphics[scale=0.7]{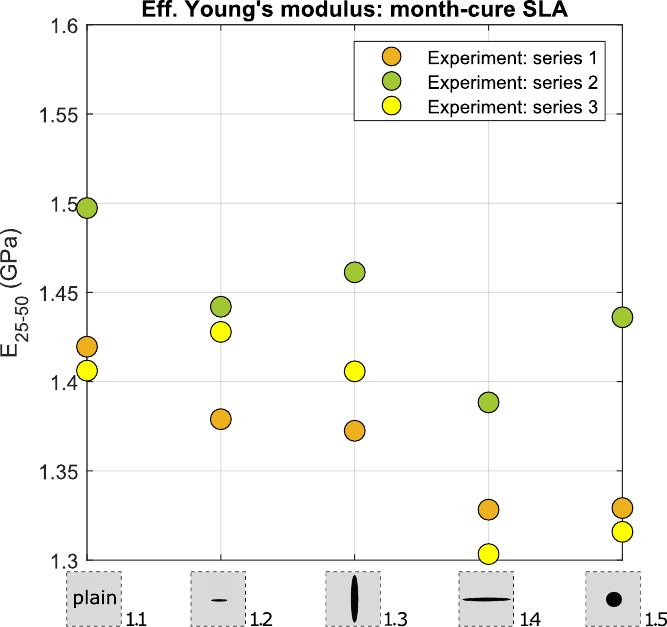}
\vspace{0.25cm}
\end{subfigure}
\qquad
\begin{subfigure}{.45\textwidth}
\includegraphics[scale=0.7]{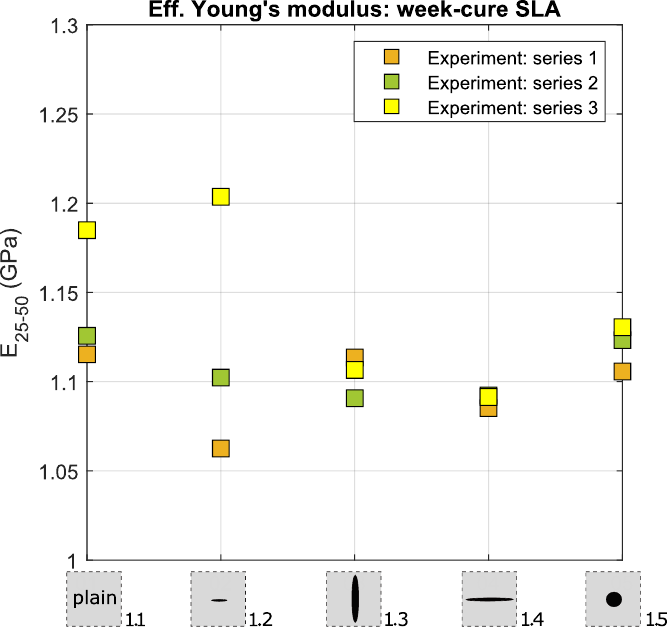}
\vspace{0.25cm}
\end{subfigure}
\begin{subfigure}{.45\textwidth}
\includegraphics[scale=0.7]{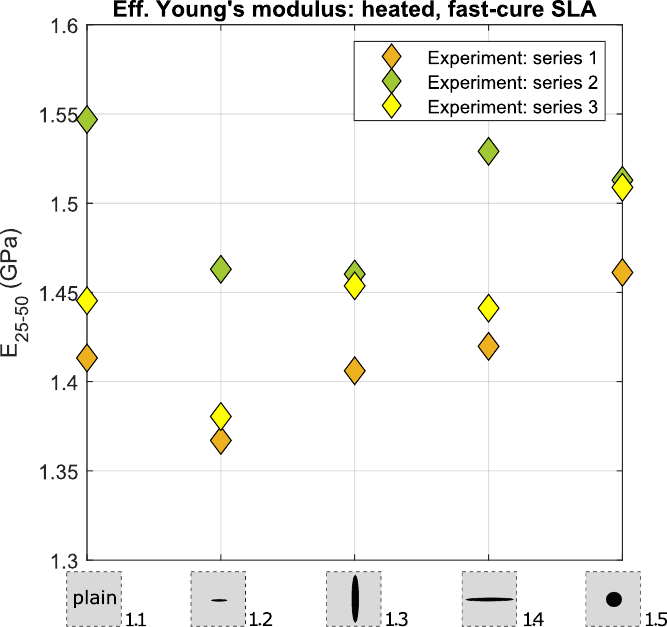}
\end{subfigure}
\qquad
\begin{subfigure}{.45\textwidth}
\includegraphics[scale=0.7]{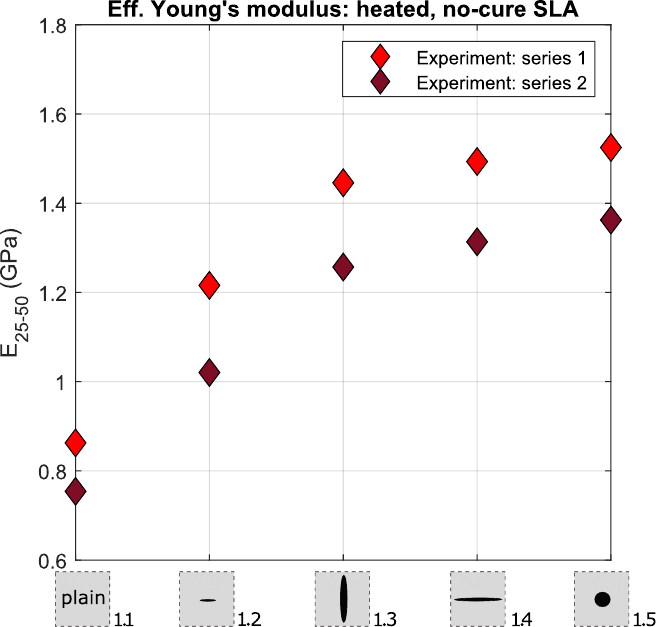}
\end{subfigure}
\caption{\footnotesize{Effective Young's moduli obtained from compression tests of SLA samples. They were printed in a series of five samples 1.1---1.5. Cores were cured for a month, week, were heated with a short cure (between 4 hours and 2 days) or almost no cure.  The results show that series differ from each other, and particular printing sessions may affect the stiffness of samples.
}}
\label{slacuringmethods}
\end{figure}

Samples 1.1---1.5 were printed using the SLA method and treated with the curing procedures described in Section \ref{stressstrainsec} (1 month at room temperature, 1 week at room temperature, 1 hour in $60^{\circ}$C chamber). Three series of samples 1.1-1.5 were printed, treated, and tested for each curing method. Two additional series of samples were tested immediately after heating (no curing time). This workflow was followed to evaluate the repeatability of results for printed samples following the same post-printing processing. Calculations of $\rm{E}_{25-50}$ for the different series processed using the same method, for each curing procedure, are displayed in Figure \ref{slacuringmethods}. The Young's modulus ($\rm{E}_{25-50}$) axis ranges are set purposely to allow visualisation of the scatter of results between different series. The scatter in $\rm{E}_{25-50}$ is overall relatively low, and below 0.2 $\rm{GPa}$. However, there is a consistent trend in the magnitude of Young's modulus between series. For example, series 2 of 1 month-cured samples have consistently higher $\rm{E}_{25-50}$ magnitudes for all 1.1---1.5 samples with respect to series 1 and 3, which were subjected to equal post-printing procedures. Similar trends are observed for all curing methods, indicating a clear influence of printing batch on the resulting mechanical properties. This observation highlights the difficulty of repeatability for SLA printed samples, even after following the same post-processing procedures.    

\begin{figure}[!htbp]
\centering
\includegraphics[scale=0.7]{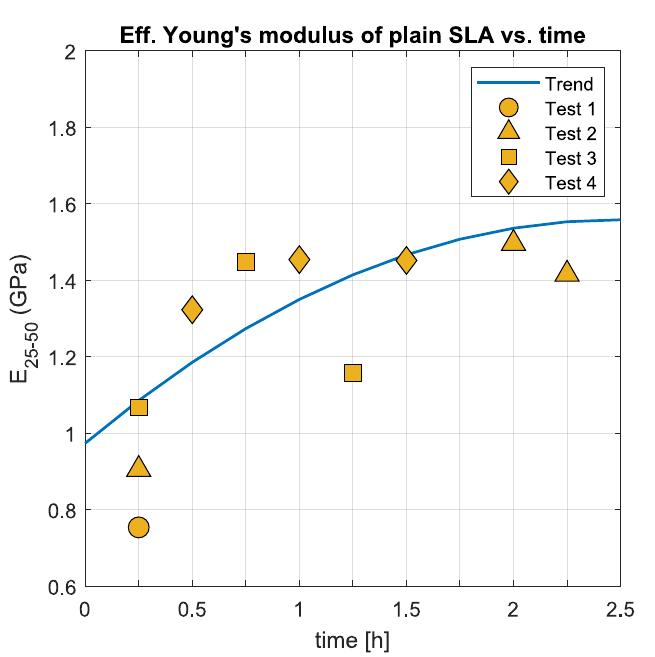}
\caption{\footnotesize{Effect of a cure time of heated samples. Decreasing temperature leads to the stiffening of the samples.}}
\label{slacuretime}
\end{figure}

The effect of the elapsed time between the curing of whole SLA samples and the execution of the uniaxial test to determine the mechanical properties of SLA samples is displayed in Figure \ref{slacuretime} for samples that were printed, cured in a $60^{\circ}$C chamber, and tested after 15 minutes to 2.5 hours. It can be seen that the elapsed time between heating and conducting the deformation experiments correlates positively with stiffness, i.e., allowing the samples to cool has a hardening effect on SLA printed samples. %
\subsection{Elastic and inelastic ranges for FDM and SLA samples}

\begin{figure}[!htbp]
\centering
\begin{subfigure}{.45\textwidth}
\includegraphics[scale=0.7]{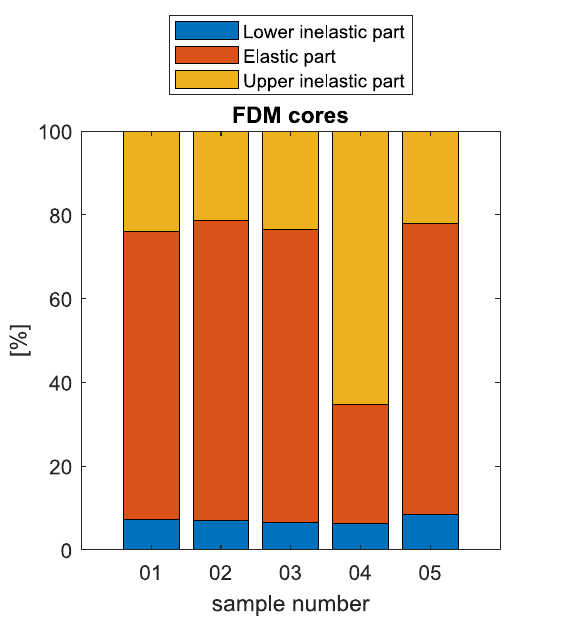}
\end{subfigure}
\begin{subfigure}{.45\textwidth}
\includegraphics[scale=0.7]{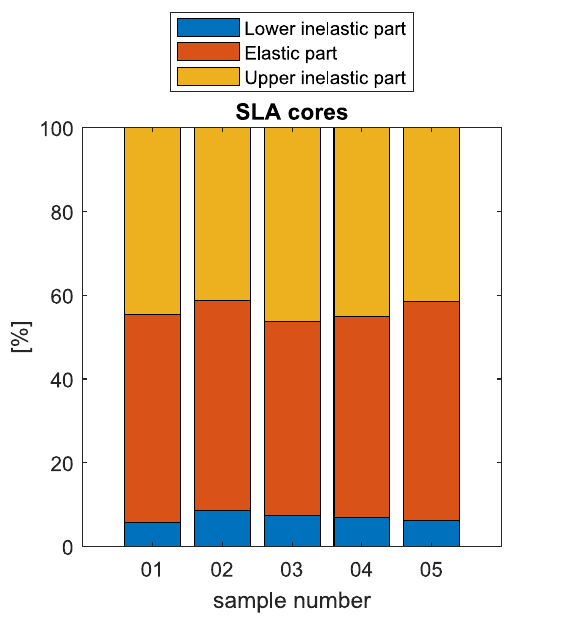}
\end{subfigure}
\caption{\footnotesize{Elastic and inelastic parts (in percentages) of an average curve of each FDM and SLA sample type, where the curve ends at the yield strength.  In the case of FDM, the yield is determined based on the flat part of the curve. In the case of SLA, yield is difficult to determine; hence, it is assumed to correspond to 6\% of strain. For FDM, on average, the elastic part takes 61.7\% of the curve.  For SLA, on average, the elastic part takes 49.3\% of the curve.}}
\label{ElasticRanges}
\end{figure}

\begin{figure}[!htbp]
\centering
\includegraphics[scale=0.7]{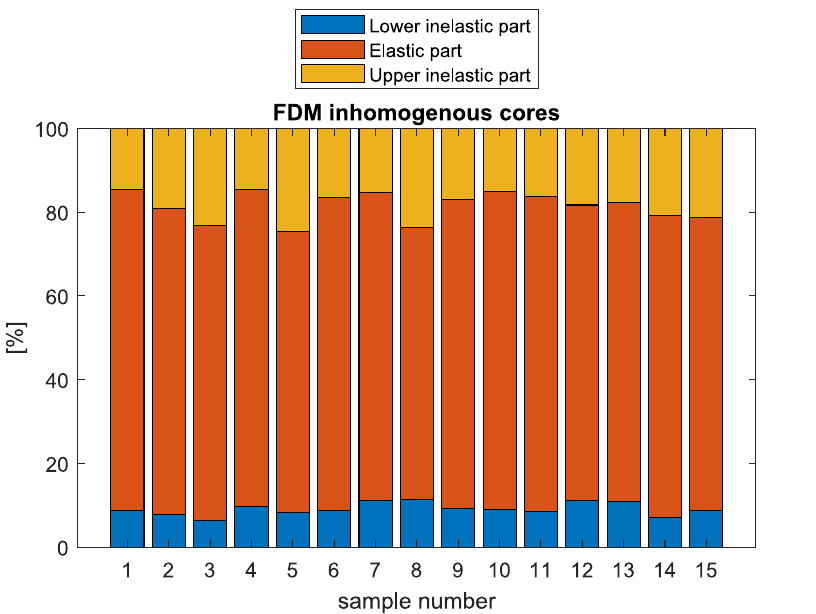}
\caption{\footnotesize{Elastic and inelastic parts (in percentages) of an average curve of each FDM secondary phase sample, where the curve ends at the yield strength. On average, the elastic part takes 72.4\% of the curve.}}
\label{ElasticRangesPhase2}
\end{figure}

Figure \ref{ElasticRanges} displays results from the strain-curves from tests on preliminary phase FDM and SLA printed samples split into the average percentages of the lower inelastic phase (origin to lower elastic limit), elastic phase (lower to upper elastic limits) and the upper inelastic phase (upper elastic limit to yield point). FDM samples deform elastically over a wider range of stress, with an average of 61.7\% of the stress-strain curves. This average is significantly decreased by sample 1.4, which displays a much smaller elastic range due to the different deformation mechanism driven by the pore space collapse. Average elastic ranges for the secondary phase of FDM printed samples (Figure\ref{ElasticRangesPhase2}) are likely more representative of the mechanical behavior of FDM samples, as none of them displays pore space collapse. They display an average elastic range of 72.4\% of the total stress-strain curve.

\par
By contrast, SLA printed samples experience a consistent, albeit smaller, elastic range of 49.3\%. The microstructural configuration of sample 1.4 does not produce deformation due to pore space collapse in SLA printed samples. However, it is important to note that the yield point for SLA printed samples is less well defined. To avoid arbitrary definitions, the yield point of SLA samples was assumed to correspond to 6\% of axial strain, throughout. 


\subsection{Prediction range from EMT using plain samples results}\label{sec:edge}
\begin{figure}[!htbp]
\centering
\includegraphics[scale=0.7]{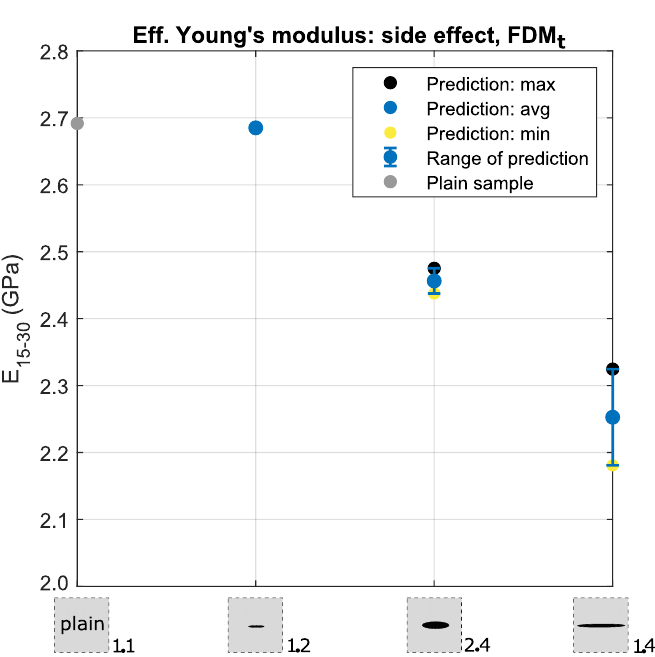}
\caption{\footnotesize{Ranges of predictions for FDM samples based on background Young's modulus, $E_0=2.69$ GPa, and background Poisson's ratio, $\nu_0=0.41$. The range of prediction increase with the augmentation of the radius of the designed crack, and with the decrease of the aspect ratio of the ellipsoid.}}
\label{fig:edge_effect}
\end{figure}

Now that we have confidently determined the elastic properties of plain SLA and FDM printed samples (section \ref{meanmodulus}), we can use these results to compute the predicted effective modulus for each sample with voids. However, EMT predictions may not be unequivocal; as they depend on the boundary conditions chosen for theoretical calculations (uniform stress or uniform strain). As a result, for certain pore geometries, different boundary conditions can result in different predicted magnitudes for equal void configurations. In such cases, the prediction of effective Young's modulus from EMT is a range, and not a single magnitude. Herein, we describe the effect that pore space volume and shape have on the range of predictions by EMT.  
Theoretically, the larger the softening effect caused by a void, the larger the range of estimation is-- meaning that the prediction becomes equivocal and, hence, less accurate.  
Both the softening effect and the prediction range are affected by the interplay of the void's orientation, shape, and size.
For the sake of analysis, let us assume a single, horizontal, oblate void, such as those embedded in samples 1.2, 1.4 and 2.4, within a FDM printed sample with two perimeters (${\rm{FDM}}_{\rm{t}}$, $E_0=2.69$ GPa, and Poisson's ratio of $\nu_0=0.41$). If we consider a fixed shape of the horizontal spheroidal void and we increase its radius (e.g., samples 1.2 and 1.4), then the increased volume of the void leads to higher inaccuracy of EMT prediction (Figure~\ref{fig:edge_effect}).
By contrast, in the case of a fixed volume of the horizontal void (e.g., samples 2.4 and 1.4), the EMT inaccuracy is affected by the aspect ratio of the spheroid; the lower the aspect ratio, the higher the inaccuracy.

\subsection{Comparison with predictions from Effective Medium Theory}

\begin{figure}[!htbp]
\centering
\begin{subfigure}{.45\textwidth}
\includegraphics[scale=0.7]{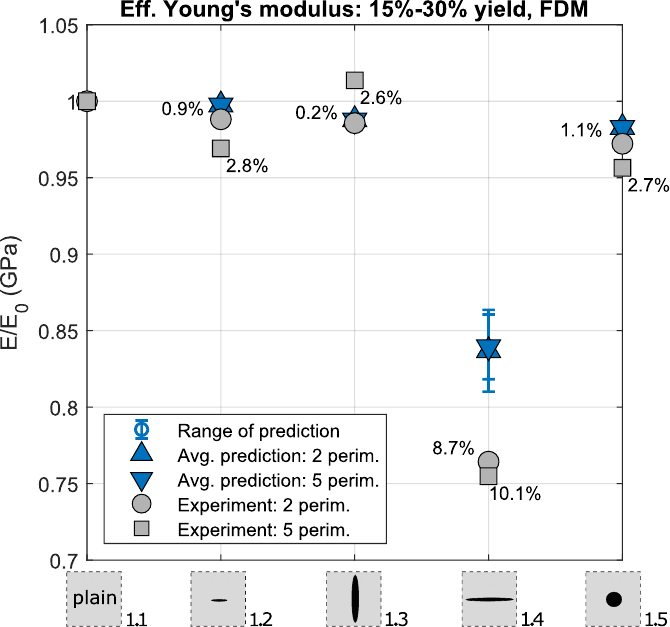}
\vspace{0.25cm}
\end{subfigure}
\qquad
\begin{subfigure}{.45\textwidth}
\includegraphics[scale=0.7]{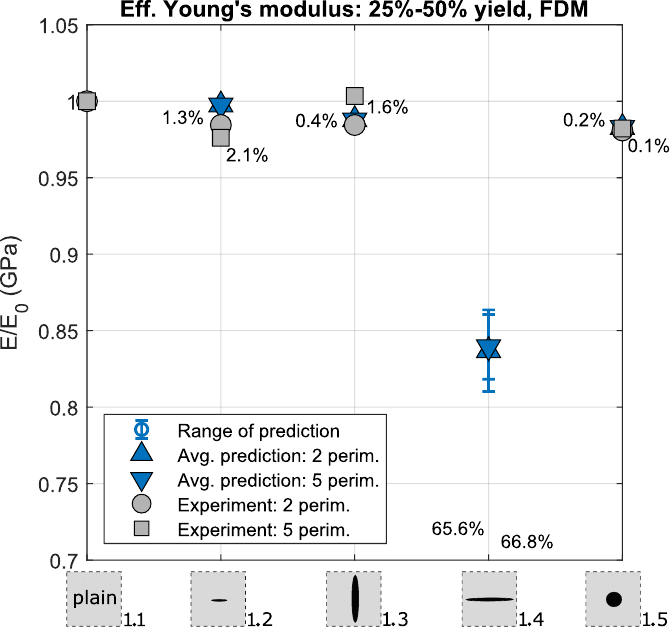}
\vspace{0.25cm}
\end{subfigure}
\begin{subfigure}{.45\textwidth}
\includegraphics[scale=0.7]{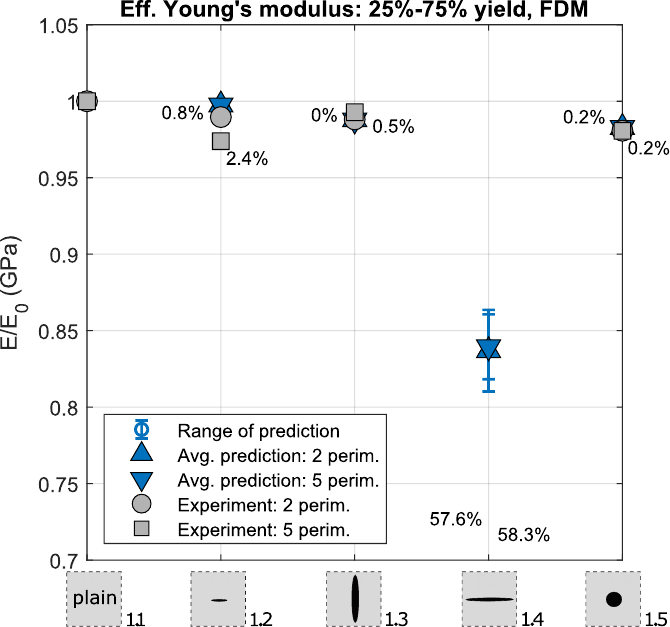}
\end{subfigure}
\qquad
\begin{subfigure}{.45\textwidth}
\includegraphics[scale=0.7]{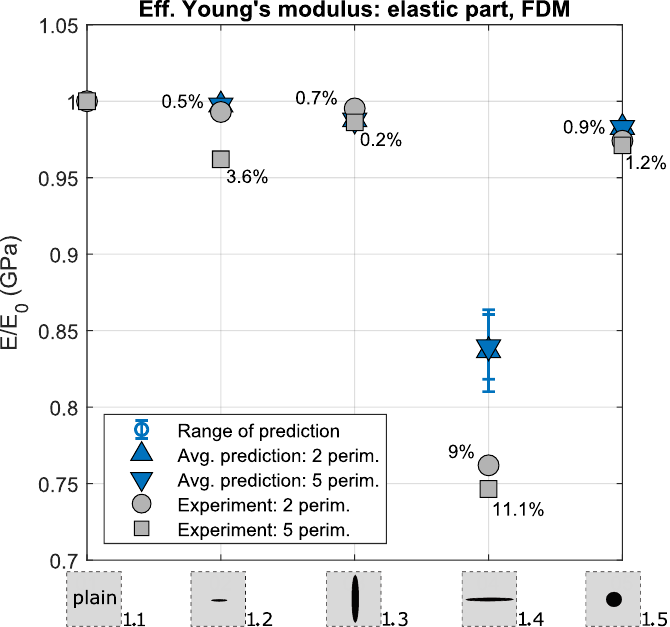}
\end{subfigure}
\caption{\footnotesize{Effective Young's moduli obtained from compression tests of FDM samples (arithmetic mean of results from six tested prints per sample type). Results are normalised by Young's modulus of the average plain core taken from Table~\ref{tab:distr}. Samples have either two or five perimeters. Each figure corresponds to different measures of Young's moduli. The relative error [\%] between mean EMT predictions and mean uniaxial results are presented. The best fit, on average $2.73\%$, appears for Young's moduli measured within the 15\%-30\% yield range and for samples having two perimeters.
}}
\label{EMTcomparisonsFDMPhase1}
\end{figure}

\begin{figure}[!htbp]
\centering
\begin{subfigure}{.45\textwidth}
\includegraphics[scale=0.7]{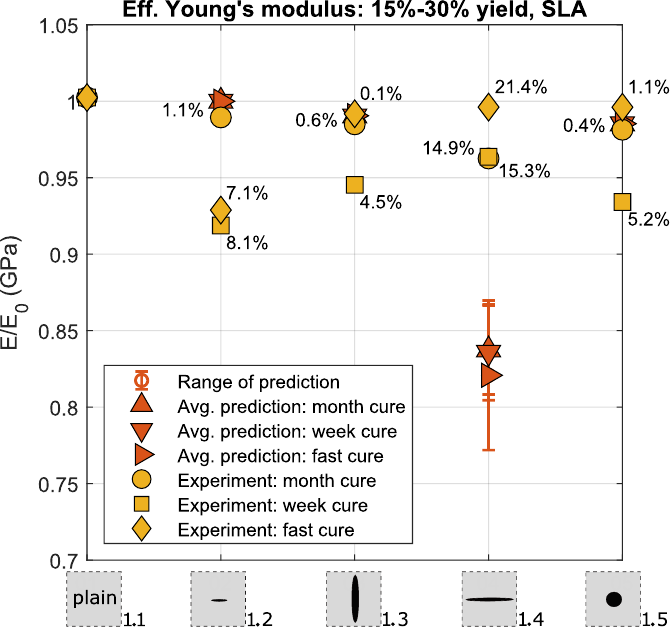}
\vspace{0.25cm}
\end{subfigure}
\qquad
\begin{subfigure}{.45\textwidth}
\includegraphics[scale=0.7]{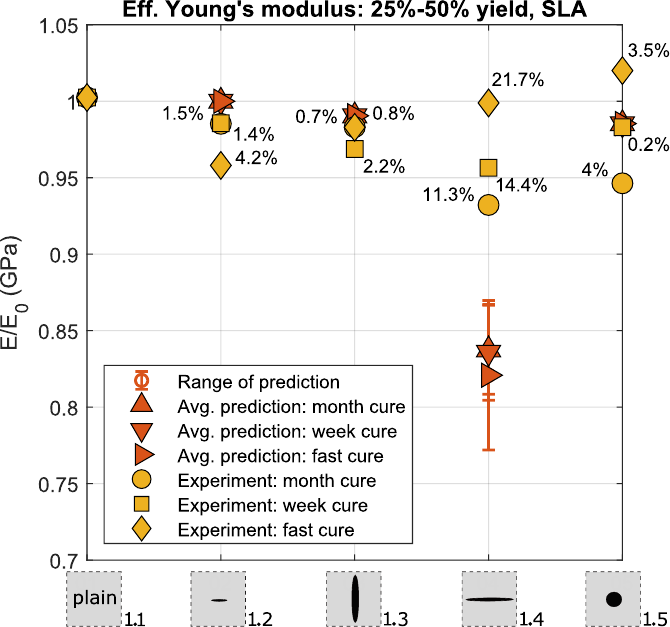}
\vspace{0.25cm}
\end{subfigure}
\begin{subfigure}{.45\textwidth}
\includegraphics[scale=0.7]{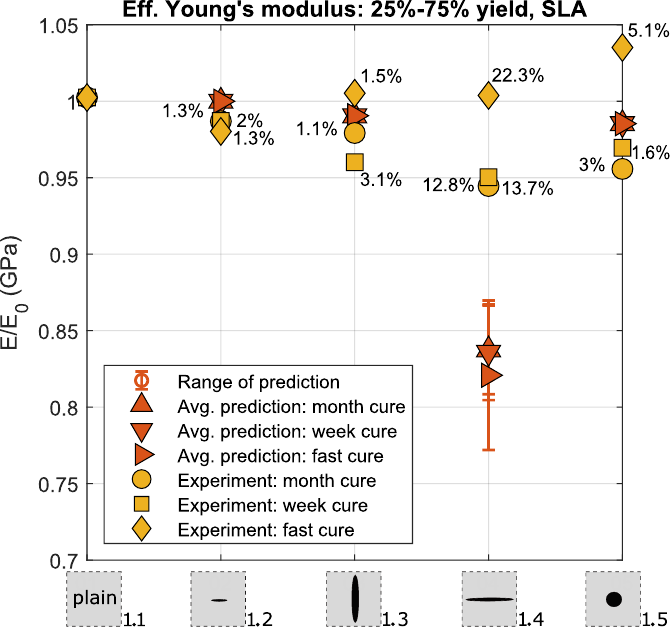}
\end{subfigure}
\qquad
\begin{subfigure}{.45\textwidth}
\includegraphics[scale=0.7]{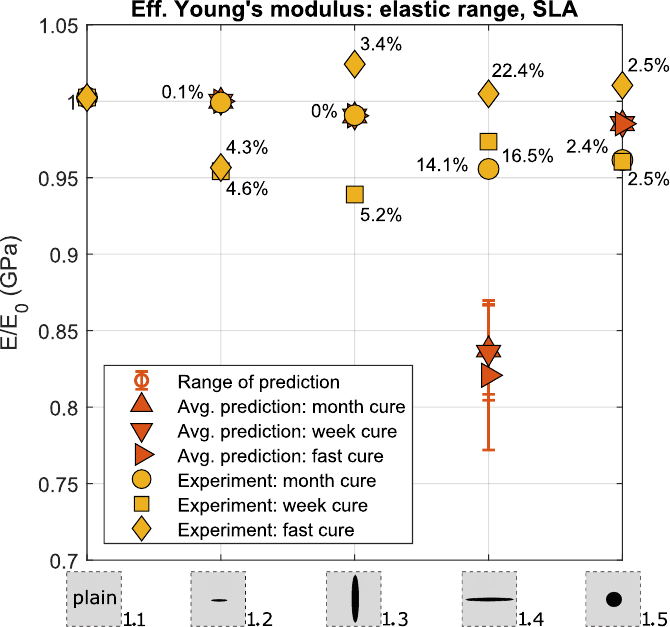}
\end{subfigure}
\caption{\footnotesize{Effective Young's moduli obtained from compression tests of SLA samples (arithmetic mean of results from three tested prints per sample type). Results are normalised by Young's modulus of the average plain core taken from Table~\ref{tab:starters_sla_young}. Samples were cured for a month (m), week (w), or were heated (h) with a short cure (up to two days). Each figure corresponds to different measures of Young's moduli. The relative differences [\%] between mean EMT predictions and mean uniaxial results are presented. 
The best fit, on average $4.07\%$, appears for Young's moduli measured within the elastic range and for samples being cured for a month.
}}
\label{EMTcomparisonsSLAPhase1}
\end{figure}

\begin{figure}[!htbp]
\centering
\begin{subfigure}{.45\textwidth}
\includegraphics[scale=0.7]{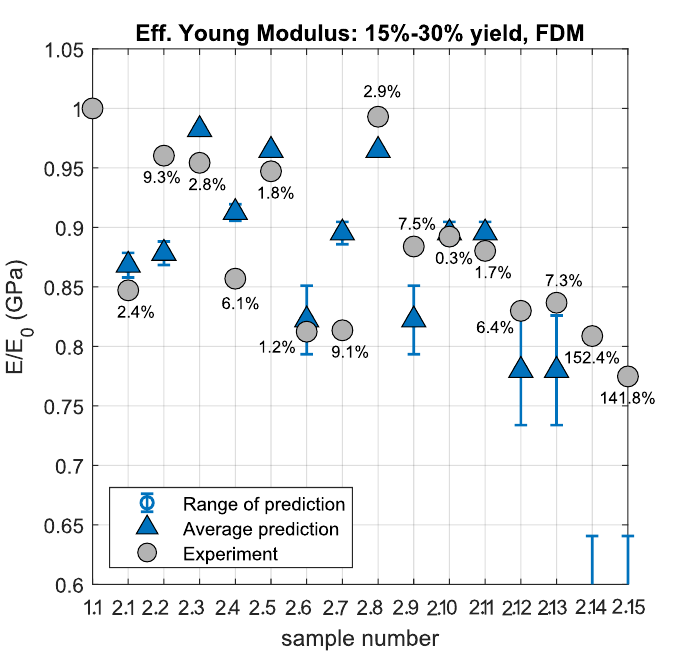}
\end{subfigure}
\qquad
\begin{subfigure}{.45\textwidth}
\includegraphics[scale=0.7]{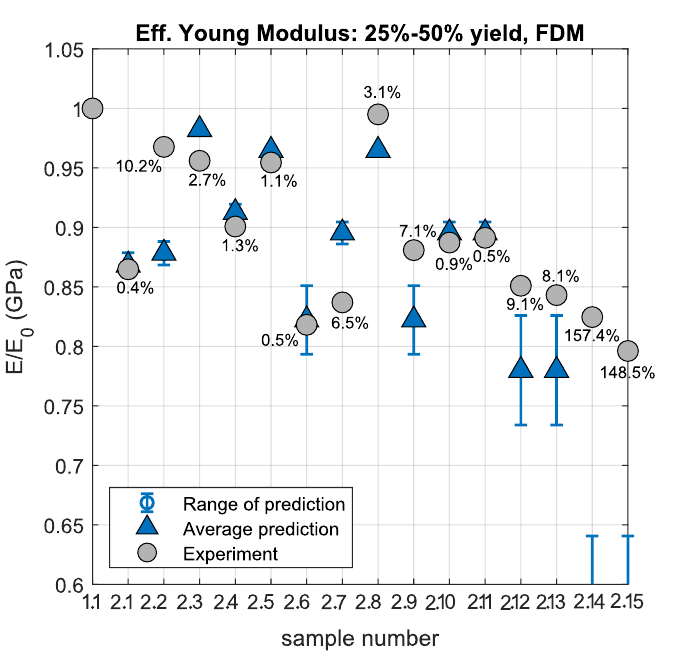}
\end{subfigure}
\begin{subfigure}{.45\textwidth}
\includegraphics[scale=0.7]{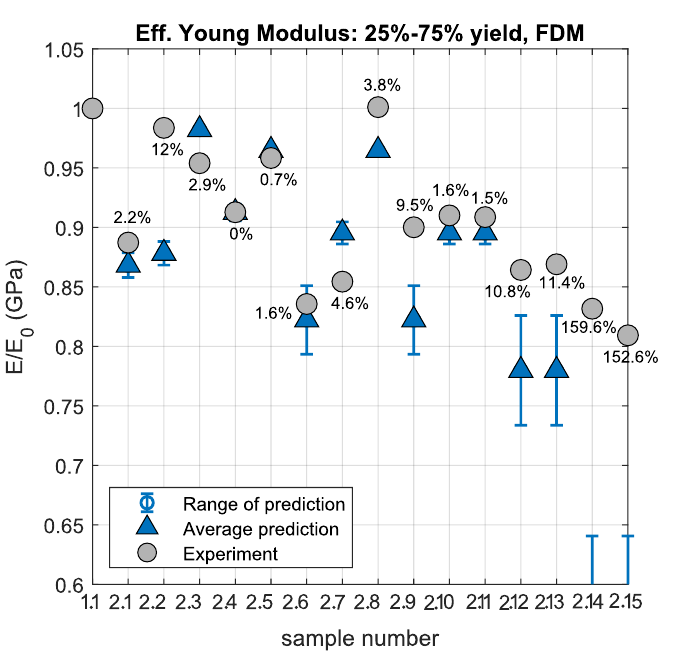}
\end{subfigure}
\qquad
\begin{subfigure}{.45\textwidth}
\includegraphics[scale=0.7]{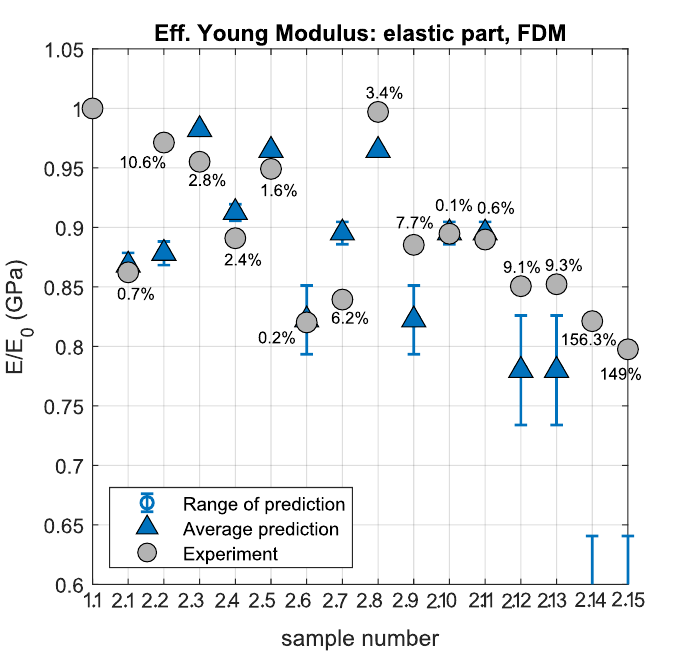}
\end{subfigure}
\caption{\footnotesize{Effective Young's moduli obtained from compressions of FDM plain samples (arithmetic mean of results from twelve tested prints) and fifteen types of FDM cores with designed voids (arithmetic mean of results from six tested prints per sample type). All samples have two perimeters. Each figure corresponds to different measures of Young's moduli. The relative error [\%] between mean EMT predictions and mean uniaxial results are presented. The best fit, on average $23.6\%$, appears for Young's moduli measured within the 15\%-30\% yield range.}}
\label{EMTcomparisonsFDMPhase2}
\end{figure}

The measured effective elastic modulus of 3D printed samples, considering four different ranges for elasticity modulus computation ($E_{15-30}$, $E_{25-50}$, $E_{25-75}$ and the whole elastic portion of the curves), are plotted along with the predicted range for effective Young's modulus calculated from EMT for each microstructral configuration for a) FDM, preliminary phase printed samples (Figure \ref{EMTcomparisonsFDMPhase1}), b) SLA, preliminary phase printed samples (Figure \ref{EMTcomparisonsSLAPhase1}), and c) secondary phase, FDM printed samples (Figure \ref{EMTcomparisonsFDMPhase2}). Details of the plotted values and the computed error of experimental results with respect to EMT predictions are listed in Tables \ref{tab:starters_prusa_young}, \ref{tab:starters_sla_young} and \ref{tab:HOT16_young} in the Appendix. The error between measured and predicted values was calculated as follows:
\begin{equation}
|R|=\frac{|E_{\rm{predicted}}-E_{\rm{measured}}|}{|E_{\rm{predicted}}|}\times 100\%\,
\end{equation}
where $E_{\rm{measured}}$ is the average effective Young's modulus (averaged over a number of repetitions), and $E_{\rm{predicted}}$ is the middle value within the EMT prediction range.

\par
The results of effective Young's modulus for FDM secondary phase samples correspond to the arithmetic mean of six tested samples per sample type, and are plotted normalised with respect to the measured Young's modulus of plain cores. The best fit to EMT predictions is achieved for effective moduli computed with $E_{15-30}$, and for samples printed with two perimeters, which display very small errors of 0.94\%, 0.23\%, and 1.08\% for samples 1.2, 1.3, and 1.5, respectively. The mismatch between measurements and predictions is highest for sample 1.4 (deformed due to pore space collapse) for any printing set-up and modulus calculation. Measured effective modulus for samples 1.4 are between 5.95\% and 65.6\% lower than EMT predictions.  
\par
The details of computed effective Young's modulus and comparison with EMT predictions for SLA printed samples are displayed in Table \ref{tab:starters_sla_young} and Figure \ref{EMTcomparisonsSLAPhase1}. Results are shown for each curing method (1 month cure ${\rm{SLA}}_{\rm{m}}$, 1 week cure ${\rm{SLA}}_{\rm{w}}$, fast and heated cure ${\rm{SLA}}_{\rm{f}}$) and for different portions of the elastic curve used for modulus computation. The closest results in terms of matching EMT predictions are achieved with 1 month cured samples (${\rm{SLA}}_{\rm{m}}$). The misfit between measurements and predictions are 1.06\%, 0.55\% and 0.39\% for samples 1.2, 1.3, and 1.5, respectively, using $E_{15-30}$ from the stress vs strain curves. Higher errors are consistently observed for sample 1.4, with a mismatch ranging between 11.3\% and 22.4\%. In contrast to FDM samples, EMT predictions underestimate the effective modulus for this microstructural arrangement with respect to measurements on SLA printed samples.
\par
Comparisons between EMT prediction with respect to measured effective moduli for secondary phase FDM printed samples (Figure \ref{EMTcomparisonsFDMPhase2}, Table \ref{tab:phase_two}) reveal that, as previously observed, lower mismatches between measurements and predictions are obtained using the $E_{15-30}$ range of the stress vs strain curve, although results are not significantly affected by the modulus computation method. Overall, the best fits to predictions (error below 5\%) are obtained for samples with simple microstructural arrangements (either one or two pores) and with lower pore volume fraction (below 3\%), which is the case for samples 2.3---2.11. The effective Young's modulus for sample 2.2 (prolate pore, 5.12\% porosity) is consistently underestimated by EMT predictions by 9-10\%. By contrast, EMT overestimates the effective modulus of sample 7 (1 oblate plus 1 prolate pore) with respect to measurements, by 4-9\%. The effective modulus for two overlapping oblate pores (sample 9) is underestimated by EMT predictions by 7-9.5\%.  
\par
More complex pore arrangements such as those constituting the internal structure of samples 2.12---2.15 display the highest errors between measurements and predictions. The higher pore fraction volume of these samples causes a larger softening effect, which results in a wider prediction range from EMT. Effective modulus measurements for samples 2.12 and 2.13 ($\phi\approx5.6\%$) lie 6-11.4\% above the upper limit of the effective modulus range predicted by EMT. By contrast, measured moduli for samples 2.14 and 2.15 ($\phi\approx11.2\%$) lie approximately 150\% above the upper bound of the predicted range.

%

\newpage
\section{Discussion}
We have produced and subjected to uniaxial compression over two hundred 3D printed samples using two different printing technologies, considering distinct printing specifications and post-printing processing methods, and twenty different microstructural configurations. This thorough approach allows us to confidently state the strengths and limitations of 3D printing to emulate solids with voids and the capacity of effective medium theory to predict their effective elastic properties. We first discuss the methods and practices that we have seen to optimize the fit of measured elastic moduli with EMT predictions. We then examine the constraints on sample dimension and microstructural design that need to be considered for manufacturing representative porous media. 

\subsection{3D printing methods and practices for best fit to EMT predictions}

The best fit between measured and predicted elasticity moduli for FDM printed samples (error less than 5\%, Table \ref{tab:starters_prusa_young}) were obtained by specifying two contour perimeters, and by computing Young's modulus using 15-30\% of the stress/strain curve. Our results suggest that the inherent anisotropy in the background medium produced by the printing method (depositional layering, raster angle), the inter-layer porosity, and the limited resolution of the printing method, does not impact significantly on the capacity of FDM samples to emulate a solid medium with voids, when sample design considerations are taken into account (see the following section). 
\par
SLA printed samples can also be used to produce controlled porous media. The best fit of measured effective elastic properties with respect to EMT predictions (error below 1\%, Table \ref{tab:starters_sla_young}) is achieved by curing the printed samples for extended periods (one month in our case), and by considering $E_{15-30}$ (as above). Successful results using SLA printing methods to match EMT predictions were also obtained by \cite{zerhouni2019numerically}. Importantly, though, our results indicate that repeatability using SLA printed samples is more difficult to achieve. The mechanical properties of SLA samples appear to be affected by uncontrolled factors (e.g. ambient temperature) that cause slightly different results between printed batches, even if subjected to equal post-printing processing. In consequence, meaningful comparisons between elastic properties of SLA cores should always be done on samples printed in the same batch. 
\par
We can conclude that both FDM and SLA methods can be used to emulate representative solids with a controlled void space whose effective elastic properties can be predicted by EMT. However, the need for long curing times of SLA samples, along with the difficulty of repeatability, added to the higher cost of this technology and the impossibility of recycling its wasted products, makes FDM a more convenient and equally effective method for this purpose.

\subsection{Considerations for microstructural design: pore volume fraction and edge effect}

In order to produce 3D printed samples of an homogeneous medium with voids that can be significantly compared to predictions from EMT, some considerations need to be taken into account. First, the pore volume fraction needs to be sufficient to induce a measurable softening effect, but, at the same time, too large pore volume fraction results in an equivocal, wide range of estimation by EMT that fails to predict the measured effective moduli of samples. 
\par
We observe that, for a pore fraction of $\phi\approx5.6\%$, such as that prescribed for samples 2.12 and 2.13, the range of effective moduli estimations from EMT becomes too wide, and lie below the measured effective properties of 3D printed samples. Results become extremely inaccurate for a pore volume fraction of $\phi\approx11.2\%$ (samples 2.14, 2.15) with errors above 150\% (Table \ref{tab:HOT16_young}). 
\par
More detailed comparisons can be made between samples with a single pore, equal shape and orientation, but different pore volume fractions. Samples 1.3 and 2.2 both contain a single, prolate, vertical pore, but with pore volume fractions of $\phi=0.46\%$ and $\phi=5.12\%$. The misfit between measurements and EMT predictions were 0.23\% and 9.34\%, respectively. 
The effect of increased pore volume is smaller when pores are spherical. Sample 1.5 ($\phi=0.87\%$) displayed a prediction error of 1\% whereas sample 2.1 ($\phi=7\%$) had a 2.4\% error with respect to predictions. Samples 1.2, 1.4 and 2.4 all contain a single, horizontal, oblate pore but with volume fractions of $\phi=0.02\%$, $\phi=1.4\%$, $\phi=1.4\%$, the latter with an aspect ratio of 0.2 as opposed to 0.1. The best fit to predictions was obtained for the smaller pore (error of 0.94\%), while samples 1.4 and 2.4 displayed errors of 8\% and 6\%, respectively. 
\par
From these results, it is tempting to conclude that smaller pore volume fractions allow EMT to better predict measured values. However, very small pores induce a negligible softening effect on samples. In these cases (e.g. sample 1.2, $\phi=0.02\%$) the measured and predicted effective Young's modulus lie within the range of measurements of plain sample modulus, which makes this comparison meaningless. A pore volume fraction of above $\approx1\%$ induces a measurable softening effect, and is therefore interesting for EMT evaluation purposes. We obtained errors of less than 5\% with respect to predictions for samples with either $\phi=1.4\%$ or $\phi=2.8\%$ (samples 2.3, 2.5, 2.6, 2.8, 2.10, 2.11) and for $\phi=5.2\%$, if the pore is spherical (sample 2.1). 
\par
Interestingly, a good fit between measurements and predictions was also obtained for overlapping pores (e.g. samples 2.8, 2.10, 2.11), in which the NIA is non-valid. Overall, experimental results show that closer interactions augment measured effective elasticity as compared to samples where voids are not overlapping, resulting in EMT predictions underestimating the effective properties of samples with interacting pores.
\par
It is also relevant to note that although the volume of the oblate, horizontal pores in samples 1.4 and 2.4 are the same, the increased aspect ratio of the pore in sample 2.4 (0.2) with respect to sample 1.4 (0.1) reduced the distance between void boundary and sample edge by 1 mm, which was enough to avoid the edge effect and a deformation mechanism dominated by pore space collapse. A distance of $\approx4$ mm from pore boundary to the sample edge is therefore recommended to avert pore collapse. 
\par
Finally, the mean error between predicted and measured effective moduli considering samples 1-13 of the secondary printing phase lie between 2-5\% (Table \ref{tab:HOT16_young}, column 16). This averaged error increases to up to 25\% when also considering samples 2.14, 2.15 (Table \ref{tab:HOT16_young}, column 17). These results highlight the strengths of EMT for predicting the effective properties of solids containing large, and even interacting, voids, as long as the pore volume fraction lies below $\approx5\%$.

\subsection{Future perspectives: pore-fluid injection into 3D printed samples for undrained poroelasticity tests}

When porous, elastic materials saturated with fluids are subjected to stress changes that may reduce or increase the volume of pore space, the fluids within them experience a change in pore fluid pressure, which can significantly influence the strength and deformation of materials \citep{Biot41,Biot55}. The most significant poroelastic effects occur in 'undrained' conditions, that is, when fluids are trapped within a poorly connected pore space and cannot immediately flow in response to stress changes. The undrained poroelastic response of a material subjected to a change in mean stress can be predicted using the relation \citep{skempton1954pore,rice1976some}:

\begin{equation}
\Delta p_{u}=B\Delta \sigma_{m}
\end{equation}

where $\Delta p_{u}$ is the change in pore fluid pressure, $\Delta \sigma_{m}$ the change in mean stress, and $B$ is the Skempton's coefficient, a scalar for isotropic materials and a tensor for anisotropic solids and/or with an anisotropic pore space \citep{wang2000theory}. Obtaining direct measurements of $B$ in the laboratory is a challenging feat, even for the 'simple' isotropic case. As a result, the effect of pore space fabric on the poroelastic response of materials has been only predicted by means of theoretical simulations \citep{Sayers95}. 

3D printed samples with voids would enable the injection of pore fluids into their isolated pores and to conduct compression experiments while measuring \textit{in-situ} pore fluid pressure change, using methods adapted from \citet{BrantutAben21} and/or \citet{proctor2020direct}. This workflow would lead to direct quantification of poroelastic parameters, which added to the absolute control of pore space fabric that 3D printing methods offer, would allow to fully describe the effect of pore geometry anisotropy on the poroelastic response of materials. This overarching goal constitutes the major motivation for the present study, and is currently a work in progress.



\section{Conclusions}
We have fabricated over two hundred 3D printed samples with voids of different sizes and geometries, using two printing technologies and various specifications, and compared their measured effective properties with predictions from EMT. From this large amount of data, we can conclude that both FDM and SLA printing technologies can emulate homogeneous solids with voids whose effective elastic properties can be predicted by EMT with an error of less than 5\%, if certain specifications and considerations are taken into account:
\begin{itemize}
\item{Best results for FDM printed samples are obtained by specifying two contour printing perimeters, and when 15 to 30\% of the stress/strain curve is considered for effective elastic modulus computation.}
\item{Best results for SLA printed samples are obtained by longer, post-printing,  curing times (one month in our case).}
\item{Although the accuracy obtained for SLA and FDM samples can be similar, the lower cost of FDM, added to the necessity of long curing times plus identified issues with repeatability and recyclability for SLA printed samples, makes FDM overall more convenient for the manufacture of solids with voids.}
\item{The pore volume fraction of the designed porous samples needs to be above 1\% to induce a measurable softening effect. However, a pore volume fraction above 5\% results in equivocal, wide range of estimation by EMT that fails to predict the measured effective moduli of samples.}
\item{Better fits are obtained for spherical pores with respect to horizontally and vertically oriented pores.}
\item{Close interactions between voids leads to a diminished softening effect that is thus overestimated by EMT predictions.}
\item{A minimum distance of 16\% of the sample diameter from the pore boundary to the sample edge should be prescribed for FDM printed samples to avoid an inelastic deformation mechanism through pore space collapse.}
\end{itemize}

\section*{Acknowledgements}
This research was supported financially by the NERC grant: ``Quantifying the Anisotropy of Poroelasticity in Stressed Rock'', NE/N007826/1 and NE/T00780X/1. We would like to thank Vitalik Ivanov, Chirs Harbord and Tongzhang Qu for their help in various laboratory tasks.









\appendix
\addcontentsline{toc}{section}{Appendices}
\section{Tables}\label{ap2}

\begin{sidewaystable}[!htbp]
\scalebox{0.85}{%
\begin{tabular}{|l|c|c|c|c|c|c|c|c|c|c|c|c|}
  \cline{2-13}
\multicolumn{1}{c|}{}& & & & & & & & & & & &       \\[-0.13in] 
 \multicolumn{1}{c|}{} &  \multicolumn{1}{|c|}{${1.1}_t$}  &  \multicolumn{1}{|c|}{${1.2}_t$}  & ${1.3}_t$ & ${1.4}_t$ & ${1.5}_t$ & ${1.1}_t-{1.5}_t$ & ${1.1}_v$ & ${1.2}_v$ & ${1.3}_v$ &${1.4}_v$ &${1.5}_v$ &${1.1}_v-{1.5}_v$   \\[+0.01in] \cline{1-13}
& & & & & & & & & & & &      \\ [-0.05in]  
exp. $E_{15-30}$&$2.69$&$2.66$ &$2.65$ &$2.06$ &$2.62$ &$2.54$ & $2.64$&$2.56$ & $2.68$& $1.99$& $2.53$& $2.48$        \\ [+0.1in] 

exp. span\,\,& $0.31$&$0.26$ &$0.29$ &$0.23$ &$0.18$ &
$-$ & $0.33$&$0.35$ & $0.30$& $0.14$& $0.14$& 
$-$  \\ [+0.1in] 

th. $E_{15-30}$& $-$&$2.69$ &$2.66$ &$2.25$ &$2.65$ &$2.60$ & $-$&$2.63$ & $2.61$& $2.22$& $2.60$& $2.51$   \\ [+0.1in] 

th. range\,\,& $-$&$\pm 0.00$ &$\pm 0.00$ &$\pm 0.07$ &$\pm 0.00$ &$-$ & $-$&$\pm 0.00$ & $\pm 0.00$& $\pm 0.06$& $\pm 0.00$& $-$  \\[+0.1in] 

${|R|}^{\rm{m}}_{15-30}$ [\%]&$-$&$0.94$ &$0.23$ &$8.66$ &$1.08$ &$2.73$ & $-$ & $2.84$& $2.62$& $10.1$& $2.68$ &$4.55$  \\ [+0.1in] 

${|R|}^{\rm{cls}}_{15-30}$ [\%]& $-$&$0.94$ &$0.22$ &$5.65$ &$1.07$ &$1.97$ & $-$&$2.84$ & $2.62$& $7.74$& $2.66$& $3.97$ \\[+0.1in]   \hline

& & & & & & & & & & & &      \\ [-0.05in]  
exp. $E_{25-50}$&$2.73$&$2.69$ &$2.69$ &$0.79$ &$2.68$ &$2.31$ & $2.68$&$2.62$ & $2.69$& $0.75$& $2.64$& $2.28$  \\ [+0.1in]  

exp. span\,\,& $0.19$&$0.19$ &$0.21$ &$0.11$ &$0.16$ &
 $-$ & $0.31$&$0.32$ & $0.16$& $0.11$& $0.15$& 
$-$        \\  [+0.1in] 

th. $E_{25-50}$&$-$&$2.72$ &$2.69$ &$2.28$ &$2.68$ &$2.60$ & $-$&$2.68$ & $2.65$& $2.25$& $2.64$& $2.55$   \\  [+0.1in] 

th. range\,\,& $-$&$\pm 0.00$ &$\pm 0.00$ &$\pm 0.07$ &$\pm 0.00$ &$-$ & $-$&$\pm 0.00$ & $\pm 0.00$& $\pm 0.06$& $\pm 0.00$& $-$         \\ [+0.1in] 

${|R|}^{\rm{m}}_{25-50}$ [\%]&$-$&$1.31$ &$0.35$ &$65.6$ &$0.20$ &$16.9$ & $-$ & $2.14$ & $1.59$ & $66.8$& $0.06$& $17.6$   \\  [+0.1in] 

${|R|}^{\rm{cls}}_{25-50}$ [\%]& $-$&$1.31$ &$0.34$ &$64.4$ &$0.19$ &$16.6$ 
& $-$&$2.14$ & $1.58$& $65.9$& $0.04$& $17.4$ \\ [+0.1in]  \hline  

& & & & & & & & & & & &      \\ [-0.05in]  
exp. $E_{25-75}$& $2.64$&$2.61$ &$2.61$ &$0.94$ &$2.59$ &$2.27$ & $2.61$&$2.54$ & $2.59$& $0.91$& $2.56$& $2.24$       \\ [+0.1in] 

exp. span\,\,& $0.23$&$0.10$ &$0.18$ &$0.07$ &$0.10$ & 
$-$ & $0.32$&$0.23$ & $0.20$& $0.10$& $0.11$& 
$-$  \\ [+0.1in] 

th. $E_{25-75}$& $-$&$2.63$ &$2.60$ &$2.21$ &$2.59$ &$2.51$ & $-$&$2.60$ & $2.58$& $2.19$& $2.56$& $2.48$   \\ [+0.1in] 

th. range\,\,& $-$&$\pm 0.00$ &$\pm 0.00$ &$\pm 0.07$ &$\pm 0.00$ &$-$ & $-$&$\pm 0.00$ & $\pm 0.00$& $\pm 0.06$& $\pm 0.00$& $-$  \\[+0.1in] 

${|R|}^{\rm{m}}_{25-75}$ [\%]&$-$&$0.79$ &$0.02$ &$57.6$ &$0.22$ &$14.7$ & $-$&$2.36$ & $0.52$& $58.3$& $0.21$& $15.4$   \\ [+0.1in] 

${|R|}^{\rm{cls}}_{25-75}$ [\%]& $-$&$0.79$ &$0.01$ &$56.2$ &$0.20$ &$14.3$ & $-$&$2.36$ & $0.51$& $57.2$& $0.20$ & $15.1$ \\[+0.1in]   \hline  

& & & & & & & & & & & &      \\ [-0.05in]  
exp. $E_{\rm{el}}$& $2.63$&$2.61$ &$2.61$ &$2.00$ &$2.56$ &$2.48$ & $2.59$&$2.49$ & $2.56$& $1.94$& $2.52$& $2.42$    \\ [+0.1in] 

exp. span\,\,& $0.17$&$0.17$ &$0.24$ &$0.33$ &$0.15$ & 
$-$ & $0.29$&$0.28$ & $0.19$& $0.25$& $0.24$& 
$-$       \\ [+0.1in] 

th. $E_{\rm{el}}$& $-$&$2.62$ &$2.60$ &$2.20$ &$2.58$ &$2.50$ & $-$&$2.59$ & $2.56$& $2.18$& $2.55$& $2.47$   \\[+0.1in] 
 
th. range\,\,& $-$&$\pm 0.00$ &$\pm 0.00$ &$\pm 0.07$ &$\pm 0.00$ &$-$ & $-$&$\pm 0.00$ & $\pm 0.00$& $\pm 0.06$& $\pm 0.00$& $-$    \\[+0.1in] 

${|R|}^{\rm{m}}_{\rm{el}}$ [\%]&$-$&$0.46$ &$0.75$ &$8.95$ &$0.88$ &$2.76$ & $-$&$3.55$ & $0.15$& $11.1$& $1.18$& $3.99$   \\ [+0.1in] 

${|R|}^{\rm{cls}}_{\rm{el}}$ [\%]& $-$&$0.46$ &$0.74$ &$5.95$ &$0.86$ &$2.00$ & $-$&$3.55$ & $0.14$& $8.76$& $1.17$& $3.40$\\ [+0.1in]  \hline  
\end{tabular}
}
\caption{\footnotesize{Mean experimental (exp.) and theoretical (th.) Young's moduli for five FDM sample types, along with the span of the experimental results, range of theoretical predictions, and the relative errors. Samples have either two (subscript $t$) or five perimeters (subscript $v$). Predictions are made based on mean Young's moduli and Poisson's ratios of plain samples (${1.1}_t$ or ${1.1}_v$). ${|R|}^{\rm{m}}_{Y}$ stands for the relative error of the mean prediction (central value within the predicted Young's modulus range) with respect to the mean measurement, ${|R|}^{\rm{cls}}_{Y}$ denotes the relative error of the closest prediction (value within the predicted Young's modulus range) with respect to the mean measurement, where $Y$ is replaced by $15-30$, $25-50$, $25-75$, or `el' indicating Young's moduli calculation method.}}
\label{tab:starters_prusa_young}
\end{sidewaystable}

\begin{sidewaystable}[!htbp]
\scalebox{0.76}{%
\begin{tabular}{|l|c|c|c|c|c|c|c|c|c|c|c|c|c|c|c|c|c|c|}
  \cline{2-19}
\multicolumn{1}{c|}{}& & & & & & & & & & & & & & & &  & &           \\[-0.13in] 
 \multicolumn{1}{c|}{} &  \multicolumn{1}{|c|}{${1.1}_m$}  &  \multicolumn{1}{|c|}{${1.2}_m$}  & ${1.3}_m$ & ${1.4}_m$ & ${1.5}_m$ & ${1.1}_m-{1.5}_m$ &${1.1}_w$ &${1.2}_w$ &${1.3}_w$ &${1.4}_w$ &${1.5}_w$ &${1.1}_w-{1.5}_w$ &${1.1}_f$ & ${1.2}_f$ &${1.3}_f$ &${1.4}_f$& ${1.5}_f$ & ${1.1}_f-{1.5}_f$   \\[+0.01in] \cline{1-19}
& & & & & & & & & & & & & & & &  &  &         \\ [-0.05in]  
  
exp. $E_{15-30}$&$1.45$&$1.43$ &$1.42$ &$1.39$ &$1.42$ &$1.42$ & $1.20$&$1.10$ & $1.14$& $1.16$& $1.12$& $1.15$& $1.52$& $1.41$& $1.51$&  $1.51$& $1.51$  & $1.49$    \\ [+0.1in] 

exp. span\,\,& $0.07$ & $0.06$ &$0.10$ & $0.04$& $0.15$& $-$& $0.12$&$0.29$ &$0.06$ &$0.09$ &$0.08$ & 
$-$ & $0.11$& $0.06$& $0.12$&  $0.11$& $0.03$ & 
$-$ \\ [+0.1in] 

th. $E_{15-30}$& $-$&$1.44$ &$1.43$ &$1.21$ &$1.42$ &$1.38$ & $-$&$1.20$ & $1.19$& $1.00$& $1.18$& $1.15$& $-$& $1.52$& $1.51$&  $1.25$& $1.50$ & $1.44$   \\ [+0.1in] 

th. range\,\,& $-$&$\pm 0.00$ &$\pm 0.00$ &$\pm 0.04$ &$\pm 0.00$ &$-$ & $-$&$\pm 0.00$ & $\pm 0.00$& $\pm 0.04$& $\pm 0.00$& $-$& $-$& $\pm 0.00$& $\pm 0.00$&  $\pm 0.07$& $\pm 0.00$ & $-$    \\[+0.1in] 

${|R|}^{\rm{m}}_{15-30}$ [\%]&$-$&$1.06$ &$0.55$ &$14.9$ &$0.39$ &$4.24$ & $-$&$8.13$ & $4.54$& $15.3$& $5.19$& $8.29$& $-$& $7.11$& $0.13$&  $21.4$& $1.09$ & $7.42$  \\ [+0.1in] 

${|R|}^{\rm{cls}}_{15-30}$ [\%]& $-$&$1.06$ &$0.55$ &$11.1$ &$0.37$ &$3.27$ & $-$&$8.13$ & $4.53$& $11.1$& $5.17$& $7.24$& $-$& $7.11$& $0.12$&  $14.5$& $1.07$ & $5.71$ \\[+0.1in]   \hline

& & & & & & & & & & & & & & & &  &   &        \\ [-0.05in]  
exp. $E_{25-50}$&$1.44$&$1.42$ &$1.41$ &$1.34$ &$1.36$ &$1.39$ & $1.14$&$1.12$ & $1.10$& $1.09$& $1.12$& $1.12$& $1.47$& $1.40$& $1.44$&  $1.46$&  $1.49$ & $1.45$ \\ [+0.1in]  

exp. span\,\,& $0.09$&$0.06$ &$0.09$ &$0.09$ &$0.12$ &$-$ & $0.07$&$0.14$ & $0.02$& $0.01$& $0.02$& 
$-$ & $0.13$& $0.10$& $0.05$&  $0.11$&  $0.05$ & 
$-$ \\  [+0.1in] 

th. $E_{25-50}$&$-$&$1.44$ &$1.42$ &$1.20$ &$1.42$ &$1.37$ & $-$&$1.14$ & $1.13$& $0.95$& $1.12$& $1.09$& $-$& $1.47$& $1.45$&  $1.20$&    $1.44$  &    $1.39$ \\  [+0.1in] 

th. range\,\,& $-$&$\pm 0.00$ &$\pm 0.00$ &$\pm 0.04$ &$\pm 0.00$ &$-$ & $-$&$\pm 0.00$ & $\pm 0.00$& $\pm 0.04$& $\pm 0.00$& $-$& $-$& $\pm 0.00$& $\pm 0.00$&  $\pm 0.07$&    $\pm 0.00$  & $-$  \\ [+0.1in] 

${|R|}^{\rm{m}}_{25-50}$ [\%]&$-$&$1.47$ &$0.74$ &$11.3$ &$3.95$ &$4.37$ &$-$ & $1.44$&$2.19$ & $14.4$& $0.23$& $4.57$& $-$& $4.20$& $0.76$& $21.7$&  $3.51$&    $7.54$    \\  [+0.1in] 

${|R|}^{\rm{cls}}_{25-50}$ [\%]& $-$&$1.47$ &$0.73$ &$7.57$ &$3.94$ &$3.43$ & $-$&$1.44$ & $2.18$& $10.3$& $0.21$& $3.53$& $-$& $4.20$& $0.75$&  $14.8$&    $3.49$ &    $5.82$ \\ [+0.1in]  \hline  

& & & & & & & & & & & & & & & &  &  &         \\ [-0.05in]
exp. $E_{25-75}$& $1.30$&$1.28$ &$1.27$ &$1.22$ &$1.24$ &$1.26$ & $1.03$&$1.02$ & $0.99$& $0.98$& $1.00$& $1.00$& $1.29$& $1.26$& $1.30$&  $1.30$& $1.34$ & $1.30$ \\ [+0.1in] 

exp. span\,\,& $0.08$&$0.05$ &$0.08$ &$0.08$ &$0.11$ &$-$ & $0.08$&$0.11$ & $0.05$& $0.03$& $0.04$& 
$-$& $0.12$& $0.10$& $0.11$&  $0.10$& $0.03$ &
$-$ \\ [+0.1in] 

th. $E_{25-75}$& $-$&$1.29$ &$1.28$ &$1.08$ &$1.28$ &$1.23$ & $-$&$1.03$ & $1.02$& $0.86$& $1.01$& $0.98$& $-$& $1.29$& $1.28$&  $1.06$ &$1.27$ &$1.22$  \\ [+0.1in] 

th. range\,\,& $-$&$\pm 0.00$ &$\pm 0.00$ &$\pm 0.04$ &$\pm 0.00$ &$-$ & $-$&$\pm 0.00$ & $\pm 0.00$& $\pm 0.03$& $\pm 0.00$& $-$& $-$& $\pm 0.00$& $\pm 0.00$&  $\pm 0.06$& $\pm 0.00$ & $-$  \\[+0.1in] 

${|R|}^{\rm{m}}_{25-75}$ [\%]&$-$&$1.31$ &$1.12$ &$12.8$ &$3.00$ &$4.56$ & $-$&$1.29$ & $3.07$& $13.7$& $1.60$& $4.91$& $-$& $1.97$& $1.48$&  $22.3$& $5.05$  &$7.70$ \\ [+0.1in] 

${|R|}^{\rm{cls}}_{25-75}$ [\%]& $-$&$1.31$ &$1.11$ &$9.01$ &$2.99$ &$3.60$ & $-$&$1.28$ & $3.06$& $9.59$& $1.58$& $3.88$& $-$& $1.97$& $1.48$&  $15.4$ &$5.03$ &$5.97$\\[+0.1in]   \hline  

& & & & & & & & & & & & & & & &  &   &        \\ [-0.05in]  
exp. $E_{\rm{el}}$& $1.39$&$1.38$ &$1.37$ &$1.32$ &$1.33$ &$1.36$ & $1.13$&$1.08$ & $1.06$& $1.10$& $1.09$& $1.09$& $1.46$& $1.39$& $1.49$&  $1.46$ &$1.47$ &$1.45$ \\ [+0.1in] 

exp. span\,\,& $0.05$&$0.06$ &$0.05$ &$0.08$ &$0.12$ &$-$ & $0.09$&$0.19$ & $0.06$& $0.02$& $0.02$& 
$-$& $0.14$& $0.11$& $0.12$&  $0.12$ &$0.09$ &
$-$    \\ [+0.1in] 

th. $E_{\rm{el}}$& $-$&$1.39$ &$1.37$ &$1.16$ &$1.36$ &$1.32$ & $-$&$1.13$ & $1.12$& $0.94$& $1.11$& $1.08$& $-$& $1.46$& $1.44$&  $1.19$ &$1.43$ &$1.38$   \\[+0.1in] 
 
th. range\,\,& $-$&$\pm 0.00$ &$\pm 0.00$ &$\pm 0.04$ &$\pm 0.00$ &$-$ & $-$&$\pm 0.00$ & $\pm 0.00$& $\pm 0.04$& $\pm 0.00$& $-$& $-$& $\pm 0.00$& $\pm 0.00$&  $\pm 0.07$& $\pm 0.00$ & $-$    \\[+0.1in] 

${|R|}^{\rm{m}}_{\rm{el}}$ [\%]&$-$&$0.08$ &$0.05$ &$14.1$ &$2.40$ &$4.16$ & $-$&$4.58$ & $5.18$& $16.5$& $2.52$& $7.19$& $-$& $4.33$& $3.40$&  $22.4$ &$2.55$ &$8.18$   \\ [+0.1in] 

${|R|}^{\rm{cls}}_{\rm{el}}$ [\%]& $-$&$0.08$ &$0.04$ &$10.3$ &$2.38$ &$3.20$ & $-$&$4.58$ & $5.16$& $12.3$& $2.50$& $6.13$& $-$& $4.33$& $3.40$&  $15.5$ &$2.53$ &$6.45$ \\ [+0.1in]  \hline  
\end{tabular}
}
\caption{\footnotesize{Mean experimental (exp.) and theoretical (th.) Young's moduli for five SLA sample types, along with the span of the experimental results, range of theoretical predictions, and the relative errors.  Samples are cured for a month (subscript $m$), a week (subscript $w$), or are fast-cured and heated (subscript $f$). Predictions are made based on mean Young's moduli and Poisson's ratios of plain samples (${1.1}_m$, ${1.1}_w$, or ${1.1}_f$). ${|R|}^{\rm{m}}_{Y}$ stands for the relative error of the mean prediction (central value within the predicted Young's modulus range) with respect to the mean measurement, ${|R|}^{\rm{cls}}_{Y}$ denotes the relative error of the closest prediction (value within the predicted Young's modulus range) with respect to the mean measurement, where $Y$ is replaced by $15-30$, $25-50$, $25-75$, or `el' indicating Young's moduli calculation method.}}
\label{tab:starters_sla_young}
\end{sidewaystable}

\begin{sidewaystable}[!htbp]
\scalebox{0.825}{%
\begin{tabular}{|l|c|c|c|c|c|c|c|c|c|c|c|c|c|c|c|c|c|}
  \cline{2-18}
\multicolumn{1}{c|}{}& & & & & & & & & & & & & & & &  &           \\[-0.13in] 
 \multicolumn{1}{c|}{} &  \multicolumn{1}{|c|}{2.1}  &  \multicolumn{1}{|c|}{2.2}  & 2.3 & 2.4 & 2.5 & 2.6 &2.7 &2.8 &2.9 &2.10 &2.11 &2.12 &2.13 & 2.14 &2.15 &2.1--2.13& 2.1--2.15   \\[+0.01in] \cline{1-18}
& & & & & & & & & & & & & & & &  &           \\ [-0.05in]  

exp. $E_{15-30}$&$2.28$&$2.58$ &$2.57$ &$2.31$ &$2.55$ &$2.19$ & $2.19$&$2.67$ & $2.38$& $2.40$& $2.37$& $2.23$& $2.25$& $2.18$& $2.09$&  $2.38$&  $2.35$        \\ [+0.1in] 

exp. span\,\,& $0.15$&$0.15$ &$0.62$ &$0.40$ &$0.58$ &$0.34$ & $0.79$&$0.34$ & $0.17$& $0.27$& $0.26$& $0.35$& $0.09$& $0.16$& $0.12$& $-$ & $-$  \\ [+0.1in] 

th. $E_{15-30}$& $2.34$&$2.36$ &$2.64$ &$2.46$ &$2.60$ &$2.21$ & $2.41$&$2.60$ & $2.21$& $2.41$& $2.41$& $2.10$& $2.10$& $0.86$& $0.86$& $2.37$& $2.17$  \\ [+0.1in] 

th. range\,\,& $\pm0.03$ &$\pm0.03$ &$\pm0.00$ &$\pm0.02$ &$\pm0.00$ & $\pm0.08$&$\pm0.03$ & $\pm0.00$& $\pm0.08$& $\pm0.03$& $\pm0.03$& $\pm0.12$& $\pm0.12$& $\pm0.86$&  $\pm0.86$& $-$ & $-$  \\[+0.1in] 

${|R|}^{\rm{m}}_{15-30}$ [\%]&$2.43$&$9.34$ &$2.84$ &$6.10$ &$1.83$ &$1.20$ & $9.13$&$2.93$ & $7.51$& $0.34$& $1.68$& $6.41$& $7.29$& $152$& $142$& $4.54$ & $23.6$ \\ [+0.1in] 

${|R|}^{\rm{cls}}_{15-30}$ [\%]& $1.26$&$8.12$ &$2.83$ &$5.37$ &$1.76$ &$0$ & $8.18$&$2.86$ & $3.86$& $0$& $0.64$& $0.48$& $1.31$& $26.2$& $20.9$&  $2.82$ & $5.58$ \\[+0.1in]   \hline

& & & & & & & & & & & & & & & &  &           \\ [-0.05in]  

exp. $E_{25-50}$&$2.36$&$2.64$ &$2.61$ &$2.46$ &$2.60$ &$2.23$ & $2.28$&$2.71$ & $2.40$& $2.42$& $2.43$& $2.32$& $2.30$& $2.25$& $2.17$&  $2.44$&  $2.41$  \\ [+0.1in]  

exp. span\,\,& $0.21$&$0.20$ &$0.64$ &$0.37$ &$0.42$ &$0.22$ & $0.82$ & $0.22$ & $0.12$& $0.19$& $0.26$& $0.26$& $0.19$& $0.17$& $0.19$ &  $-$ & $-$ \\  [+0.1in] 

th. $E_{25-50}$&$2.37$&$2.40$ &$2.68$ &$2.49$ &$2.63$ &$2.24$ & $2.44$&$2.63$ & $2.24$& $2.44$& $2.44$& $2.13$& $2.13$& $0.87$& $0.87$ & $2.40$ & $2.20$  \\  [+0.1in] 

th. range\,\,& $\pm 0.03$&$\pm 0.03$ &$\pm 0.00$ &$\pm 0.02$ &$\pm 0.00$ &$\pm 0.08$ & $\pm 0.03$ & $\pm 0.00$ & $\pm 0.08$& $\pm 0.03$& $\pm 0.03$& $\pm 0.13$& $\pm 0.13$& $\pm 0.87$& $\pm 0.87$&  $-$ & $-$  \\ [+0.1in] 

${|R|}^{\rm{m}}_{25-50}$ [\%]&$0.40$&$10.2$ &$2.67$ &$1.29$ &$1.05$ &$0.50$ & $6.52$&$3.14$ & $7.14$& $0.90$& $0.45$& $9.12$& $8.11$& $157$& $149$ & $3.96$ & $23.8$ \\ [+0.1in] 

${|R|}^{\rm{cls}}_{25-50}$ [\%]& $0$&$8.95$ &$2.65$ &$0.53$ &$0.99$ &$0$ & $5.53$&$3.07$ & $3.51$& $0$& $0$& $3.04$& $2.08$& $28.7$& $24.3$& $2.33$& $5.56$\\ [+0.1in]  \hline  

& & & & & & & & & & & & & & & &  &           \\ [-0.05in]  
exp. $E_{25-75}$& $2.34$&$2.59$ &$2.52$ &$2.41$ &$2.53$ &$2.20$ & $2.25$&$2.64$ & $2.37$& $2.40$& $2.40$& $2.28$& $2.29$& $2.19$& $2.13$&  $2.40$&    $2.37$       \\ [+0.1in] 

exp. span\,\,& $0.20$&$0.21$ &$0.66$ &$0.32$ &$0.39$ &$0.22$ & $0.77$&$0.26$ & $0.14$& $0.19$& $0.27$& $0.27$& $0.19$& $0.14$& $0.17$&  $-$&    $-$  \\ [+0.1in] 

th. $E_{25-75}$& $2.29$&$2.32$ &$2.59$ &$2.41$ &$2.54$ &$2.17$ & $2.36$&$2.54$ & $2.17$& $2.36$& $2.36$& $2.06$& $2.06$& $0.84$& $0.84$&  $2.32$&  $2.13$ \\ [+0.1in] 

th. range\,\,& $\pm 0.03$&$\pm 0.03$ &$\pm 0.00$ &$\pm 0.02$ &$\pm 0.00$ &$\pm 0.08$ & $\pm 0.02$&$\pm 0.00$ & $\pm 0.08$& $\pm 0.02$& $\pm 0.02$& $\pm 0.12$& $\pm 0.12$& $\pm 0.84$& $\pm 0.84$&  $-$& $-$  \\[+0.1in] 

${|R|}^{\rm{m}}_{25-75}$ [\%]&$2.18$&$12.0$ &$2.87$ &$0.02$ &$0.67$ &$1.63$ & $4.57$&$3.77$ & $9.47$& $1.65$& $1.49$& $10.8$& $11.4$& $160$& $153$& $4.81$& $25.0$ \\ [+0.1in] 

${|R|}^{\rm{cls}}_{25-75}$ [\%]& $0.98$&$10.8$ &$2.86$ &$0$ &$0.60$ &$0$ & $3.56$&$3.70$ & $5.76$& $0.60$& $0.44$& $4.64$& $5.22$& $29.8$& $26.3$&  $3.01$& $6.35$ \\[+0.1in]   \hline  

& & & & & & & & & & & & & & & &  &           \\ [-0.05in]  
exp. $E_{\rm{el}}$& $2.27$&$2.55$ &$2.51$ &$2.34$ &$2.49$ &$2.15$ & $2.21$&$2.62$ & $2.33$& $2.35$& $2.34$& $2.23$& $2.24$& $2.16$& $2.10$&  $2.36$&    $2.33$     \\ [+0.1in] 

exp. span\,\,& $0.24$&$0.14$ &$0.71$ &$0.38$ &$0.54$ &$0.36$ & $0.83$&$0.23$ & $0.09$& $0.19$& $0.27$& $0.23$& $0.13$& $0.18$& $0.22$&  $-$ & $-$ \\ [+0.1in] 

th. $E_{\rm{el}}$& $2.28$&$2.31$ &$2.58$ &$2.40$ &$2.53$ &$2.16$ & $2.35$&$2.53$ & $2.16$& $2.35$& $2.35$& $2.05$& $2.05$& $0.84$& $0.84$& $2.32$& $2.12$ \\[+0.1in] 
 
th. range\,\,& $\pm 0.03$&$\pm 0.03$ &$\pm 0.00$ &$\pm 0.02$ &$\pm 0.00$ &$\pm 0.08$ & $\pm 0.02$&$\pm 0.00$ & $\pm 0.08$& $\pm 0.02$& $\pm 0.02$& $\pm 0.12$& $\pm 0.12$& $\pm 0.84$& $\pm 0.84$&  $-$& $-$ \\[+0.1in] 

${|R|}^{\rm{m}}_{\rm{el}}$ [\%]&$0.69$&$10.6$ &$2.75$ &$2.37$ &$1.61$ &$0.25$ & $6.25$&$3.35$ & $7.68$& $0.07$& $0.61$& $9.07$& $9.28$& $156$& $149$& $4.20$& $24.0$ \\ [+0.1in] 

${|R|}^{\rm{cls}}_{\rm{el}}$ [\%]& $0$&$9.36$ &$2.74$ &$1.62$ &$1.55$ &$0$ & $5.26$&$3.28$ & $4.03$& $0$& $0$& $2.99$& $3.19$& $28.2$& $24.5$&  $2.62$& $5.78$ \\ [+0.1in]  \hline  
\end{tabular}
}
\caption{\footnotesize{Mean experimental (exp.) and theoretical (th.) Young's moduli for fifteen FDM sample types, along with the span of the experimental results, range of theoretical predictions, and the relative errors. Predictions are made based on mean Young's moduli and Poisson's ratios of a plain sample ${1.1}_{t}$. ${|R|}^{\rm{m}}_{Y}$ stands for the relative error of the mean prediction (central value within the predicted Young's modulus range) with respect to the mean measurement, ${|R|}^{\rm{cls}}_{Y}$ denotes the relative error of the closest prediction (value within the predicted Young's modulus range) with respect to mean measurement, where $Y$ is replaced by $15-30$, $25-50$, $25-75$, or `el' indicating Young's moduli calculation method.}}
\label{tab:HOT16_young}
\end{sidewaystable}

\newpage

\bibliographystyle{apa-good}
\bibliography{LibraryUpperCase}
\end{document}